\documentclass[aps,prx,longbibliography,twocolumn,superscriptaddress]{revtex4-2}

\usepackage{amsfonts}
\usepackage{amsmath}
\usepackage{graphicx}
\usepackage{dsfont}
\usepackage{diagbox}
\usepackage{array}
\usepackage{amssymb}
\usepackage{bbm}
\usepackage{float}
\usepackage{subfigure}
\usepackage{MnSymbol}
\usepackage{url}
\usepackage{times}
\usepackage{manfnt}
\usepackage{booktabs}
\usepackage{xcolor}
\definecolor{darkblue}{HTML}{004D6B}
\definecolor{darkred}{HTML}{8c1515}
\definecolor{darkgreen}{HTML}{006400}
\usepackage{hyperref}
\hypersetup{
    pdftitle={Learning transition of topological surface codes},
	colorlinks=true,
	urlcolor=darkred,
	citecolor=darkblue,
	linkcolor=darkred,
	breaklinks
}
\pagecolor{white}
\usepackage{ulem}
\usepackage{physics}
\usepackage{enumitem}
\usepackage{wasysym}
\usepackage{comment}
\usepackage[export]{adjustbox}

\begin{document}

\title{Learning transitions of topological surface codes}

\author{Finn Eckstein}
\affiliation{Institute for Theoretical Physics, University of Cologne, Z\"ulpicher Straße 77, 50937 Cologne, Germany}
\affiliation{The Hong Kong University of Science and Technology (Guangzhou), Nansha, Guangzhou, 511400, Guangdong, China}

\author{Bo Han}
\affiliation{Institute for Theoretical Physics, University of Cologne, Z\"ulpicher Straße 77, 50937 Cologne, Germany}
\affiliation{Department of Condensed Matter Physics, Weizmann Institute of Science, Rehovot 7610001, Israel}

\author{Simon Trebst}
\email{trebst@thp.uni-koeln.de}
\affiliation{Institute for Theoretical Physics, University of Cologne, Z\"ulpicher Straße 77, 50937 Cologne, Germany}

\author{Guo-Yi Zhu}
\email{guoyizhu@hkust-gz.edu.cn}
\affiliation{The Hong Kong University of Science and Technology (Guangzhou), Nansha, Guangzhou, 511400, Guangdong, China}

\date{\today}

\begin{abstract}
    For the surface code, topological quantum order allows one to encode logical quantum information in a robust, long-range entangled many-body quantum state. 
    However, if an observer probes this quantum state by performing measurements on the underlying qubits, 
    thereby collecting an ensemble of highly correlated classical snapshots, two closely related questions arise:    
    (i) do measurements decohere the topological order of the quantum state; 
    and (ii) how much of the logical information can one learn from the snapshots? 
    Here we address these questions for measurements in a uniform basis on all qubits. 
     We find that for generic measurement angles, sufficiently far away from the Clifford X, Y, and Z directions (such as the $X+Y+Z$ basis)
     the logical information is {\sl never lost} in one of the following two ways:
    (i) for weak measurement, the topological order is absolutely robust; 
    (ii) for projective measurement, the quantum state inevitably collapses, but the logical quantum information is faithfully transferred from the quantum system to the observer in the form of a tomographically complete classical shadow. 
    At these generic measurement angles and in the projective-measurement limit, 
    the measurement ensemble enforced by Born probabilities can be represented by a 2D tensor network  that can be fermionized 
    into a disordered, free-fermion network model in symmetry class DIII, which gives rise to a Majorana ``metal" phase. 
    When the measurement angle is biased towards the X or Z limits, a critical angle indicates the threshold of a learning transition 
    beyond which the classical shadow no longer reveals full tomographic information (but reduces to a measurement of the logical X {\sl or} Z state).
    This learning transition can be described in the language of the network model as a ``metal to insulator" transition. 
    For the anisotropic ``insulating'' angles, there is also a finite learnability threshold in terms of the measurement strength, 
    above which the surface code breaks down. In terms of universality, this latter transition is generically a Nishimori transition. 
    Our results can be generalized from {\sl qubit} measurement to {\sl syndrome} measurement, demonstrating that -- in the presence of coherent errors -- 
    even syndrome measurements can enable the observer to learn the logical information 
    and collapse the topological order across a learnability threshold. 
 \end{abstract}

\maketitle

\section{Introduction}

Topological quantum order originates from long-range entanglement~\cite{Chen2010}, which protects logical quantum information against decoherence~\cite{Preskill2002}. The realization of a topological quantum memory in the toric code~\cite{Kitaev2003} or its planar variant, the surface code~\cite{Bravyi1998}, lies at the center of the field of quantum error correction, while also serving as paradigmatic models in quantum statistical physics where their quantum topology is understood as a quantum superposition of a massive amount of loop configurations. 
These quantum memories are known to be stable against Hamiltonian perturbations~\cite{Trebst07toriccode,Tupitsyn2010, Dusuel2011}, wave function deformations~\cite{Fradkin04RK, Henley04RK, Castelnovo2008, Troyer10topocrit, haegeman2015shadows, Zhu19}, and noise~\cite{Preskill2002,Fan23toriccode, Lee23criticalityunderdecoherence, Wang23aversym, You24weaksym, Luo24weaksym, Wang24strtowksym}.  
The topological protection mechanism also prevents an observer from learning the encoded information locally~\cite{Kitaev2003}.
Such learning becomes possible only if the observer is sufficiently entangled with the physical qubits of the system~\cite{teleportcode},
the measurement strength becomes sufficiently strong~\cite{NishimoriCat}, or the readout fidelity surpasses a finite threshold~\cite{Preskill2002}.
The physics underlying such {\sl learning transitions} has attracted some recent interest~\cite{Zdeborova16inferencethreshold, Puetz25learning, Vasseur22learnability, Bao20learning, Potter24learnability,Ippoliti24learnability, You24learning, Vasseur25mixedlearning, Nahum25bayesiancriticalpoints, Vasseur25monitoredfluctuatinghydrodynamics, Kim25learning}, as its description leads to disordered statistical models which distinguish themselves 
by their {\sl correlated disorder} -- mandated by Born measurement probabilities or Bayesian inference likelihood -- and the critical theories emerging 
at the phase transitions. 
%

\begin{figure*}[th!] 
    \centering
    \includegraphics[width=\textwidth]{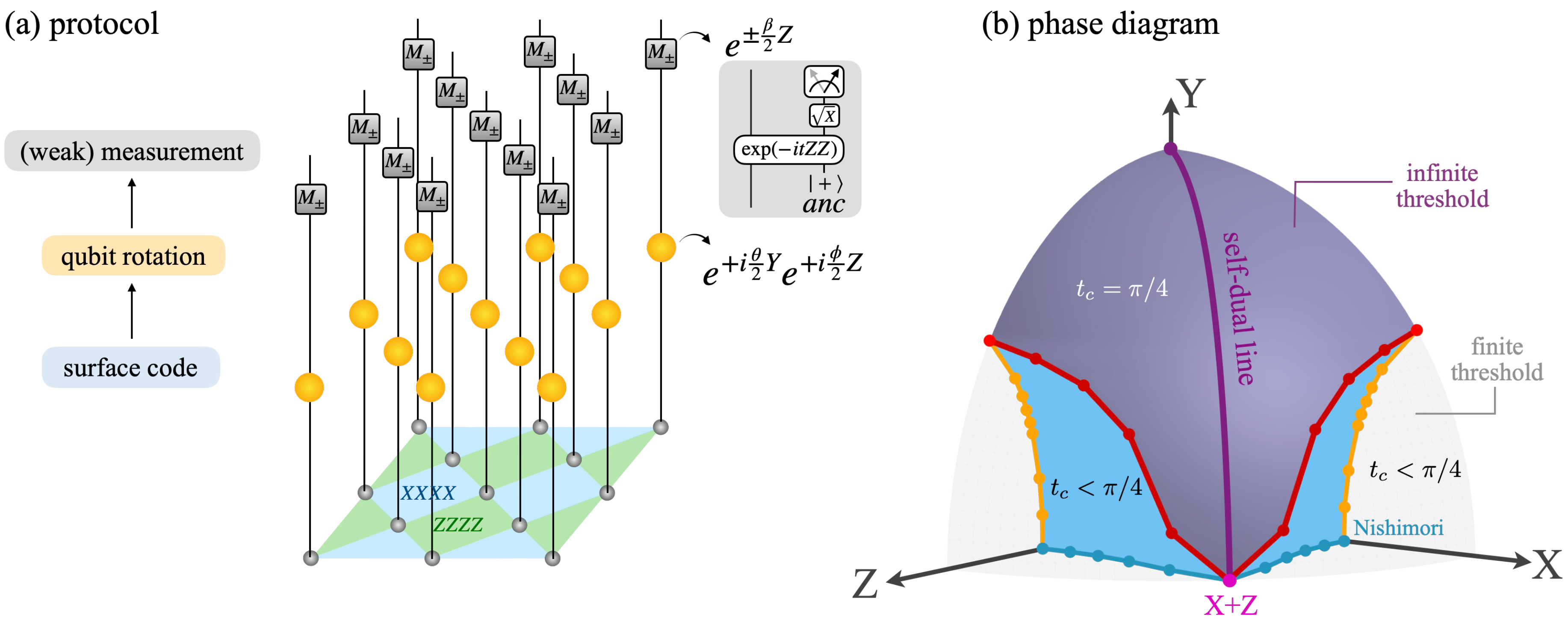}
    \caption{{\bf (a) Protocol:} A surface code is (weakly) measured in the basis $\hat{\sigma}^{\theta,\phi} \equiv \sin\theta \cos\phi X + \sin\theta \sin\phi Y + \cos\theta Z$ for all qubits. 
    The protocol consists of 3 steps: firstly, prepare a surface code in certain state in the logical space; secondly, apply a layer of coherent rotation to all qubits by $e^{i\theta/2 Y}e^{i\phi/2 Z}$; thirdly, apply a layer of weak measurement in the computation basis i.e.\ Pauli $Z$ basis. The weak measurement is implemented by introducing an ancilla prepared in $|+\rangle$ state ($X\ket{+}=\ket{+}$), and entangling it with the physical qubit by Ising interaction evolution up to time $t\in[0,\pi/4]$, and then measuring it out in the Pauli $Y$ basis. 
    As shown in Ref.~\cite{NishimoriCat}, this ancilla-assisted operation effectively implements a weak measurement gate $e^{\pm \beta Z /2 }$ with tunable measurement strength $\beta=\tanh^{-1}\sin(2t)\in[0,\infty)$. In the following, we will also refer to $t$ as the effective measurement strength. In particular, $t=\pi/4$ describes the projective-measurement limit with infinite strength $\beta=\infty$. 
    {\bf (b) Three-dimensional Born average phase diagram of parameter $(t,\theta,\phi)$:} the radial coordinate corresponds to the compactified measurement strength $t$, while the direction of the coordinate corresponds to the measurement basis of the physical qubit, akin to the Bloch sphere of a spin-$1/2$, that can be parametrized by the spherical coordinates $(\theta,\phi)$. 
    The blue surface denotes the critical boundary of surface code, which sets the learnability threshold. 
    Below the threshold $t<t_c$, the surface code remains topological, and the measurement outcomes do not extract useful information about the logical state in thermodynamic limit. 
    In the conventional measurement basis such as $X$ or $Z$ direction, a finite threshold $t_c=0.143\pi$~\cite{Puetz25percolation,teleportcode, NishimoriCat, Wan25nishimori} is established by the Nishimori transition (blue dot). The same Nishimori universality describes the whole transition line (blue dots) until the self-dual limit at $X+Z$, which is captured instead by the weak self-dual universality class~\cite{Wang25selfdual}. 
    Around the self-dual directions (purple line), $t_c=\pi/4$ ($\beta_c=\infty$), the maximal infinite threshold is found to hold for a big phase region (denoted by the purple surface) enveloped by the red doted line as the boundary. 
    }
    \label{fig:protocol}
 \end{figure*}

In this manuscript, we study the learning transitions of topological surface codes induced by measuring all physical qubits in a given basis with a tunable measurement strength, i.e.\ interpolating between weak and projective measurements.  After repeated experiments of such an all-qubit measurement, the observer is left with an ensemble of classical snapshots, while the quantum system assumes an ensemble of post-measurement states which, in combination, one can view as a hybrid mixed state~\cite{Wang25selfdual}. Such a remnant mixed state inherits certain nontrivial correlations from the original topologically ordered, long-range entangled state. For example, when the measurement basis is perfectly aligned along the $Z$ or $X$ directions, and if the measurement is projective, it is known that the quantum memory turns into a classical memory,
the quantum loop condensate is dephased into a classical loop gas ensemble~\cite{teleportcode}. 
In this case, the learning transition itself, which is driven by tuning the measurement strength, can be mapped to a Nishimori transition \cite{teleportcode,Nishimori1981,nishimori2001statistical} and, from a symmetry perspective, can be understood 
as the strong-to-weak spontaneous symmetry breaking of the underlying 1-form symmetry~\cite{Lee23criticalityunderdecoherence,Luo24weaksym, Wang23aversym, Wang24strtowksym, You24weaksym}. 
When, however, the measurement basis is rotated into a generic direction much less is known  about the learnability threshold of the surface code.
A first indication that there might be an interesting uncharted territory to be explored arose in previous work by the authors~\cite{teleportcode}
where 
it was shown that the learnability threshold is pushed from finite to {\sl infinite} 
when the measurement basis is rotated to $X+Z$, half way between the Clifford $X$ and $Z$ angles.
At this point the self-duality of the surface code (which, in the associated 2D statistical model, can be represented as a non-invertible Kramers-Wannier (KW) duality) was understood to originate this seemingly extraordinary level of robustness~\cite{Wang25selfdual}.
This naturally leads one to ask what happens as one further moves away from these Clifford angles and measures the physical qubits in a basis corresponding to an arbitrary spherical angle on the Bloch sphere. 
What is the nature of the ensuing mixed state that collects the whole ensemble of the classical snapshots? 
What can one learn about the logical qubit from such a mixed state? 
And when the measurement is weak or if measurement readout errors exist, is there (still) a learnability transition? 
These are precisely the questions that we will address in the following.

Before we provide a short summary of our main results  let us quickly mention that one can view the rotation of the measurement angle, or the weakening of the entanglement between the system and the observer, as ``coherent errors"~\cite{bravyi2018correcting, Beri23toriccodecoherenterror, Beri24coherenterror, Iverson2020, Behrends24Xcohernoise, Behrends24coheralldir, Chen25coherenterror}. Here we prefer a more neutral notation and refer to ``coherent rotations" instead as, form the learning perspective that we adopt, such coherent rotations enable varying the learnability threshold by changing the measurement basis.

\section{Summary of our results}
\label{sec:summary}

\subsubsection*{Model}

Our model consists of a 3-step protocol as illustrated in Fig.~\ref{fig:protocol}(a): 
(i) preparing a topological surface code state $\ket{\psi_\mu}$ where $\mu $ labels an arbitrary state of the single logical qubit in the Bloch sphere (e.g. $\mu=\pm$ as eigenstate of logical Pauli $X$, or $\mu=0(1)$ as eigenstate of logical Pauli $Z$) encoded in the surface code;
(ii) applying a layer of coherent rotation 
\begin{equation}
	U(\theta, \phi)  = e^{i\frac{\theta}{2}\sum_j Y_j} e^{i \frac{\phi}{2}\sum_j Z_j}
    \label{eq:Utheta}
\end{equation}
uniformly to all qubits; 
(iii) applying weak measurement of strength $\beta\in[0,\infty]$ in the computation basis, described by the Kraus operator 
\begin{equation}
	M_\mathbf{s} = e^{\frac{\beta}{2} \sum_j s_j Z_j }/\sqrt{e^\beta + e^{-\beta}} \,,
    \label{eq:M_s}
\end{equation}
where $\mathbf{s}$ labels the measurement outcome in the form of a  bit string, e.g.\ $\mathbf{s}=+-++\cdots$.
As shown inside the gray box in Fig.~\ref{fig:protocol}(a), the weak measurement can be implemented by projectively measuring an ancilla qubit which is first coupled to the physical qubit by an entangling gate with parameter $t\in [0,\pi/4]$ such that $\tanh\beta = \sin(2t)$~\cite{NishimoriCat,Chen25nishimori}. 
One can also simply view $t\in[0,\pi/4]$ as a compactified parameter for the measurement strength where  $t=\pi/4$ corresponds to the projective-measurement limit $\beta=\infty$. 

Putting the three steps together, we arrive at an ensemble of states
\begin{equation}
    \{ P(\mathbf{s}) \ ,\ \ket{\psi_\mu(\mathbf{s}) }\}\ ,
\end{equation} 
where for each classical measurement record $\mathbf{s}$, the conditional quantum state is given by the following normalized wave function
\begin{equation}
    \ket{\psi_\mu(\mathbf{s}) } = \frac{M_\mathbf{s} U(\theta, \phi) \ket{\psi_\mu}}{||M_\mathbf{s} U(\theta, \phi) \ket{\psi_\mu}||} \ ,
\end{equation} 
whose probability follows from the Born's rule
\begin{equation}
P(\mathbf{s}) = \braket{\psi_\mu(\mathbf{s})} \ .
\end{equation}
The phase diagram we discuss below is for the Born ensemble of states, or equivalently, the mixed state $\rho$ that packs the classical records and the quantum states together into a quantum-classical hybrid mixture~\cite{Wang25selfdual,Puetz25learning}
\begin{equation}
    \rho = \sum_\mathbf{s} \ketbra{\mathbf{s}} \otimes \ketbra{\psi_\mu(\mathbf{s})} \ .
\end{equation}
The task of learning is to infer the logical information $\mu$ based on the classical measurement records $\mathbf{s}$, that are sampled from the probability distribution $P(\mathbf{s})$. In a simple scenario, if the surface code is projectively measured without any coherent rotation in the $Z$ basis, one obtains the classical snapshots in the $Z$ basis, which are known to form a classical loop gas across the lattice, defined by the spin downs $Z=-1$~\cite{Wen05stringnetcondensation, Wen19science}. Such a measurement also allows one to directly readout the eigenvalue of the logical $Z_L$ operator of the surface code. 
Similarly, if the coherent rotation rotates $Z$ to the $X$ basis ($\theta=\pi/2,\ \phi=0$), the physical qubits are collapsed into the $X$ basis, where the spins with $X=-1$ forms a loop gas across the lattice, which is dual to the $Z$-loop gas~\cite{Wang25selfdual}. 
Now the question is, what happens in between? Specifically, when the qubits are collapsed at the self-dual angle $(\theta=\pi/4,\ \phi=0)$, i.e.\ in the $X+Z$ basis, the ensemble of classical snapshots has been shown to be {\sl critical} in Ref.~\cite{Wang25selfdual}, but what does that imply for the associated distribution of logical qubit snapshot? 
More generally, what happens to these distributions if the measurement angle is rotated to more general directions in the Bloch sphere and if the measurement is weak? 
%

\begin{figure*}[t!] 
    \centering
    \includegraphics[width=.99\textwidth]{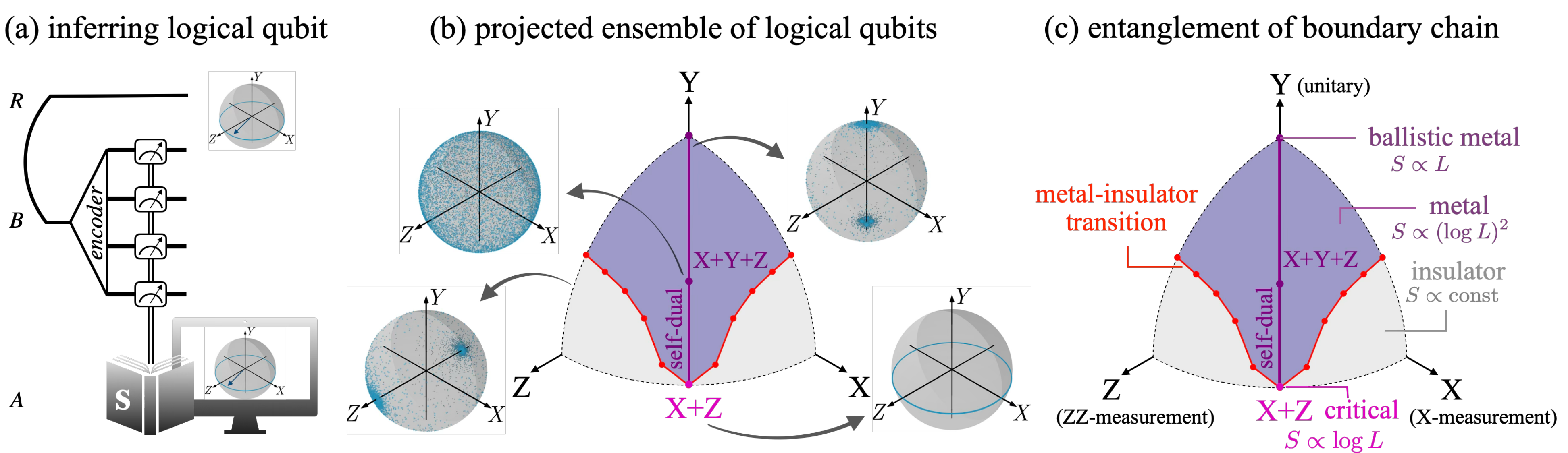}
    \caption{
    {\bf projective-measurement limit.}~  
    (a) 
    A deterministic isometry matrix encodes the logical qubit to the physical qubits of the surface code. The logical qubit is initially maximally entangled with the reference ($R$) qubit. The physical qubits are subjected to projective measurement, whose outcome is recorded by an observer ($A$). 
    For each observed measurement record $\mathbf{s}$, the observer ($A$) can decode (by inverting the unitary encoder that prepares the surface code state) the corresponding state of reference ($R$) qubit, which is a snapshot of the logical qubit. 
    These measurement records $\mathbf{s}$ form the raw data from which we construct classical shadows of the logical qubit state for tomography~\cite{HuangKuengPreskill2020shadowtomography}.~
    (b) Taking together all the possible measurement records, $\{P(\mathbf{s}), \ket{\psi_R(\mathbf{s})}\}$ forms a projected ensemble of $R$. 
    In the gapped phases (white), the ensemble converges to a bimodal distribution, revealing the logical $Z$ or logical $X$ state. 
    At the self-dual critical point $X+Z$, the ensemble delocalizes to a uniform distribution, but confined to the real $XZ$ plane, forming a ring on the equator of the Bloch sphere. 
    In the metal gapless phase (purple), the ensemble fully delocalizes to a Haar measure uniform coverage of the Bloch sphere. Nevertheless, it crossovers to the bimodal distribution at the Clifford $Y$ limit, which is a singular point. 
    The shown ensembles of states are computed for surface code of code distance $d=8$.
    (c) 
    The boundary entanglement entropy has distinct scaling at the phases. 
    In the gapped phases, it exhibits area law. 
    At self-dual critical point, critical logarithmic scaling is found. 
    Into the metal gapless phase, $\log^2 L$ dominates the entanglement entropy scaling. 
    The $Y$ limit corresponds to a ballistic metal, exhibiting volume law scaling. 
    The Y-axis is described by purely unitary evolution. By contrast, the X and Z axes are described by pure measurement evolutions. 
    The circuit and the boundary state is shown in Fig.~\ref{fig:bilayertensornetwork}(c). 
    }
    \label{fig:fig2projphasediagram}
 \end{figure*}
 
\subsubsection*{Phase diagram}

To address the above questions, we have charted out the spherical phase diagram shown in Fig.~\ref{fig:protocol}(b). 
Akin to a Bloch sphere our phase diagram is characterized by a pair of spherical measurement angles $\theta,\phi$ and a measurement strength $t$ that indicates the radial  direction.
Its color coding indicates different regimes of finite and infinite {\it learnability thresholds} which we have determined by calculating the coherent information~\cite{Nielsen96coherentinfo, Lloyd97coherentinfo, Fan23toriccode,teleportcode} as an information-theoretic order parameter of the topological order of the surface code in the presence of measurements. 
Our underlying numerical computations have charted out the various phase boundaries, most prominently the purple surface around the $Y$-direction where, akin to the self-dual $X+Z$ point discussed above, the learning transition happens only in the projective-measurement limit. 
From a mixed-state perspective, surface code remains, for these measurement angles, robust against arbitrarily strong, but non-projective measurements. 
This should be contrasted with the finite thresholds 
observed around the corners surrounding the $X$ or $Z$ measurement directions,
where already for some finite measurement strength the surface code $\{\ket{\psi(\mathbf{s})}\}$ is destroyed and useful information in $\{ P(\mathbf{s})\}$ can be extracted by the observer. Cast in terms of mixed-state terminology, the surface code is dephased for finite measurement strength, with the quantum loop gas loosing its coherence and transitioning into a classical loop gas.

\subsubsection*{Ensemble of logical states}

In the projective-measurement limit, all physical qubits are collapsed into classical states along the measurement direction.  
In tandem, the logical qubit is also collapsed resulting in a snapshot along a direction, which is determined by the isometry that maps the obtained physical many-qubit snapshot to a logical single-qubit snapshot. Such a logical qubit snapshot can also be materialized by employing a reference qubit $R$, which  in the initial step of preparing the surface code is maximally entangled with the logical qubit, see Fig.~\ref{fig:fig2projphasediagram}(a). The collapse of the physical qubits of the surface code then gives rise to a projected ensemble~\cite{Ho23projectedensemble} of the reference qubit $R$ as illustrated for various measurement angles in Fig.~\ref{fig:fig2projphasediagram}(b). 
For a conventional measurement basis such as $Z$ or $X$, the logical snapshots align along the $Z$ or $X$ axis, forming a bimodal distribution.
Rotating the measurement angle towards the self-dual angle $X+Z$, we find that the projected ensemble forms a uniform ring covering the ``equator" of the Bloch sphere, with the logical state remaining confined to the real plane. 
As we move out of this plane and towards to the $Y$-direction, we find that the projected ensemble forms a uniform coverage of the Bloch sphere 
as representatively illustrated for the $X+Y+Z$ magic angle. 
This is an important observation as, in terms of the learning problem considered here, it indicates that -- 
even though the measurement basis of the physical qubits is uniform toward a fixed angle --
the snapshot of the logical qubit enables {\sl full tomography} for the logical qubit -- without the need for randomized measurements~\cite{Zoller23randomized} or measuring for multiple angles. 
Such full tomography is possible everywhere on the purple surface, but the $Y$-direction which is again a Clifford limit 
where the projected ensemble is distorted back to a bimodal distribution. 
Notice that such a transition of the projected ensemble from a Haar-measure uniform distribution to a bimodal distribution has recently also been realized in a long-range interacting circuit~\cite{Ho25coherence}. 

\subsubsection*{Boundary entanglement entropy}

The measurement-induced mixed state can be characterized by its 2D quantum topology (such as its topological degeneracy or coherent information)
or, alternatively, by its (1+1)D boundary entanglement entropy. 
In more technical terms, the post-measurement state of the surface code can be represented, for every measurement outcome $\mathbf{s}$, by
a random 2D PEPS (projected entangled pair state) which can be mapped to a (1+1)D monitored quantum circuit dynamics~\cite{Wang25selfdual}. 
The final state of this circuit dynamics captures the complexity of the boundary state.
We are particularly interested in characterizing the latter in the projective-measurement limit (at $t=\pi/4$)
for arbitrary measurement angles as summarized in the phase diagram of Fig.~\ref{fig:fig2projphasediagram}(c).
The boundary entanglement entropy exhibits an area law in the surface code phase and the gapped dephasing phases around the $X$ and $Z$ corners, but displays logarithmic scaling $S_{\rm vN}\propto \log L$ at the critical points (pink and red dots in the phase diagram). 
For generic measurement angles, i.e.\ for angles corresponding to the purple regime in Fig.~\ref{fig:fig2projphasediagram}(c) 
the boundary entanglement entropy scales $S_{\rm vN}\propto \log^2 L$, which is reminiscent of the entanglement structure of the 
Goldstone modes in a ``metallic" phase~\cite{Jian2022criticality, Nahum23majorana, Fidkowski23majorana, KlockeBuchhold2023majorana, Jian2023criticality}.
The limiting case of measurements along the $Y$-direction is singular and exhibits volume-law scaling $S_{\rm vN}\propto L$ because the corresponding (1+1)D circuit dynamics consists exclusively of maximally entangling Majorana swap gates, which is a spacetime-dual~\footnote{The duality here refers to exchanging space and time dimension, which should be distinguished from Kramers Wannier duality~\cite{Wang25selfdual} for the whole line between $Y$ and $X+Z$ discussed in this paper} unitary circuit with ballistic growth of entanglement~\cite{Prosen20dualunitary}.

\subsubsection*{Metal-insulator transition}

In connecting the monitored (1+1)D fermion circuit in the projective-measurement limit to a free-fermion model, 
one maps the non-unitary circuit dynamics in (1+1)D dimensions to 
a unitary, non-interacting fermion model in (2+1)D spatial dimensions with a static Hermitian Hamiltonian~\cite{Jian2022criticality, Jian2023criticality}, coupled to static disorder that follows the Born's distribution. Microscopically, this is done by relating each non-unitary circuit gate, as a transfer matrix in the time dimension, to a unitary scatterer in two spatial dimensions, as in the standard Chalker-Coddington network model construction~\cite{Chalker2002, Jian2022criticality}. 
As a consequence, one can identify the entanglement structures stabilized by the (1+1)D monitored fermion circuit, 
with the ground states (and their respective entanglement structures) of a (2+1)D non-interacting free-fermion model,
i.e.\ a metallic or insulating state. 
Specifically, this connects the phase with long-range (logarithmic scaling) boundary entanglement (i.e.\ the purple phase in our 
phase diagram) to a {\sl metallic} state, which (in its free-fermion formulation) is known to exhibit such logarithmically scaling entanglement \cite{Klich2006,Wolf2006}.
In contrast, the phases near $X$ and $Z$ anisotropic limit exhibit area-law entangled (1+1)D fermion chain, 
which corresponds to a short-range correlated (2+1)D {\sl insulating} free-fermion system. 

In this language, some of the learning transitions discussed in this paper can be recast as a {\sl metal-insulator} transition.
For instance, when varying the measurement angle from a generic direction (e.g.\ around the $X+Y+Z$ point) to one of the Clifford directions 
near the $X$ and $Z$ corners, one transitions precisely from a metallic (long-range) entanglement structure to an insulating (short-range)
entanglement structure, see the phase diagram in Fig.~\ref{fig:fig2projphasediagram}(c).   
As one tracks this metal-insulator transition (red dots in our phase diagram) as one approaches the self-dual $X+Z$ point (pink dot),
the system acquires weak self-dual symmetry resulting in a modified critical point, extensively chacterized in Ref.~\cite{Wang25selfdual}.
In comparison, the generic metal-insulator transition in our phase diagram has only been partially studied along the $XY$ line --
where the qubit-measurement model considered here can be mapped to a syndrome-measurement model~\cite{Beri23toriccodecoherenterror} as discussed  in Sec.~\ref{sec:XYline} below, 
and critical exponents and conformal data remain an open question.
In this paper we discuss the boundary-entanglement scaling of this generic metal-insulator transition, 
but leave it to future work for an in-depth investigation of the criticality theory of this point. 

\subsubsection*{Symmetry class}

For our model we find that for generic measurement angles the relevant symmetry class of the free-fermion model corresponding to our circuit is symmetry class DIII. 
Specifically, the transfer matrix of the tensor network can be mapped to that of a Chalker-Coddington network model, which respects time-reversal ($T^2 = -1$), particle-hole ($C^2 = +1$), and chiral symmetries -- precisely the symmetries of symmetry class DIII~\cite{Jian2023criticality}. 
However, along three special lines in the phase diagram, i.e.\ the $ZX$~\cite{Wang25selfdual}, $XY$, and $YZ$ lines of the phase diagram in Fig.~\ref{fig:fig2projphasediagram}(c), 
the tensor network remains in symmetry class D, where the latter two are related to what has been observed for {\sl syndrome} measurements in Refs.~\cite{Beri23toriccodecoherenterror,Beri24coherenterror}.
On these special lines, the tensor network can be decomposed into two decoupled Majorana fermion network models that are time-reversal partners of each other; each copy is in symmetry class D. Such a decomposition, however, does not exist away from the three special lines.  
We will analyze the $XY$ line further in Sec.~\ref{sec:XYline} and Fig.~\ref{fig:equivsyndromemeas} and, in particular, discuss its equivalence to the error model previously discussed in Refs.~\cite{Beri23toriccodecoherenterror,Beri24coherenterror}. 

\section{Learning transition for tunable measurement strength}

When the measurement strength is tuned above a certain threshold, the observer obtains certain logical information about the surface code, and meanwhile collapses the surface code into a post-measurement projected ensemble of states -- this threshold defines our learnability transition~\cite{Puetz25learning}. 
To diagnose this learning transition in our model, we will rely on two principal observables: 
(i) the {\sl coherent information} of the 2D surface code; 
(ii) the {\sl entanglement entropy} of its 1D boundary quantum state. 
The former literally discriminates two phases: learnable and unlearnable. 
The latter characterizes the fluctuations and complexity of the remaining quantum system, which can also signal the quantum critical point.
But before we introduce these observables in the following, we will briefly review two representations of our surface code model, 
which we previously introduced in Ref.~\cite{teleportcode}, that will allow for the efficient calculation of these observables:
(i) a 2D (random) tensor network representation; (ii) a (1+1)D quantum circuit which can also be recast in the form of a fermion model.

\subsection{Tensor network and quantum circuit representation}
\label{sec:TNandQC}

\subsubsection*{Tensor network representation \& Ashkin-Teller model}

\begin{figure*}[t!]
    \centering
    \includegraphics[width=\textwidth]{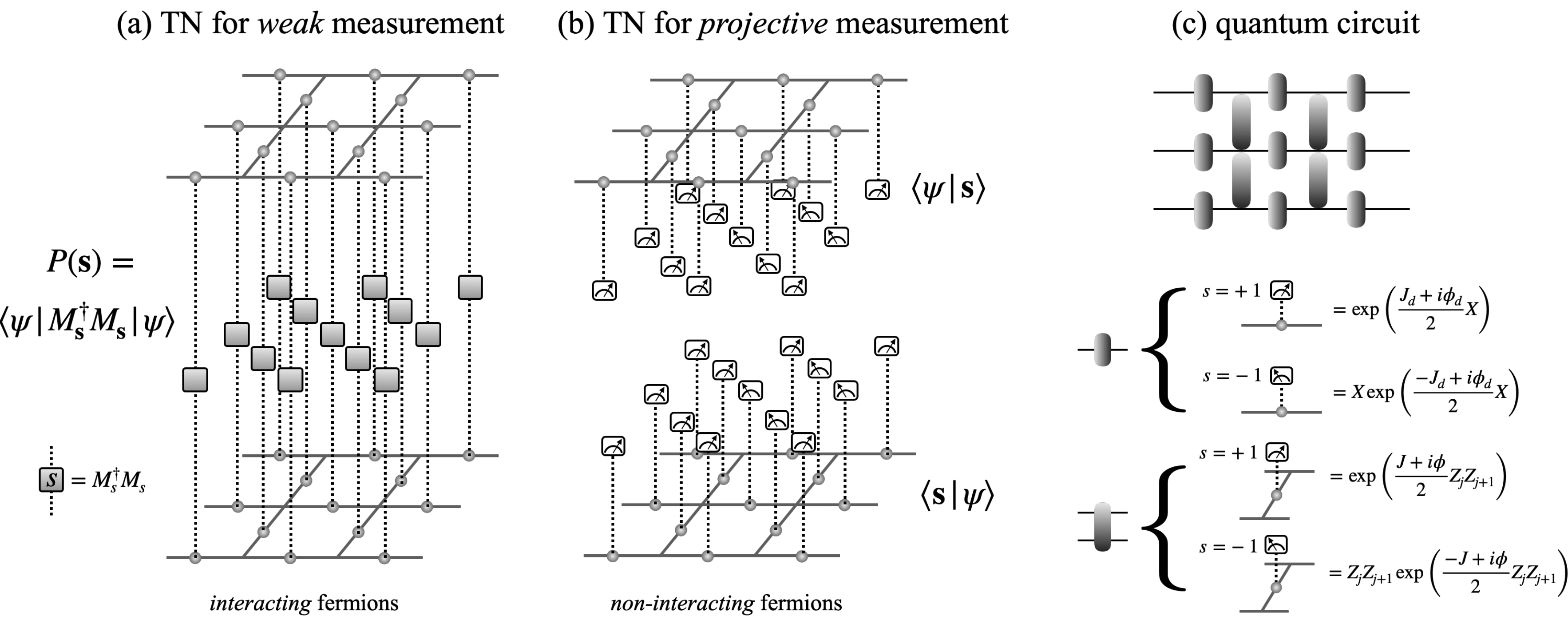}
    \caption{
        {\bf Random 2D tensor network and (1+1)D quantum circuit}.
        The top and the bottom layer of the tensor network captures the bra and ket of the quantum state as a PEPS. 
        (a) For weak measurement, the Kraus operator $M_{\mathbf{s}}^\dagger M_\mathbf{s}$ glues the two layers. 
        (b) For projective measurement, $M^\dagger_\mathbf{s} M_\mathbf{s}$ is a projector, such that $P(\mathbf{s}) = |\bra{\mathbf{s}}\ket{\psi}|^2$, which means the two layers can be factorized. As will be shown in Sec.~\ref{subsec:transfermatrix}, each layer alone corresponds to a random Ising model with complex coupling, which can be fermionized into a free fermion network model. In this aspect, weakening the measurement strength in (a) induces fermion interactions. 
        Both (a) and (b) as 2D tensor network can be interpreted as (1+1)D (non-unitary) quantum circuit that propagates a 1D quantum chain in the space direction, updated by the transfer matrix. 
    (c) The ket layer of tensor network in (b) can be identified as a monitored (1+1)D quantum circuit. The gate parameters are controlled by the coherent rotation $U(\theta, \phi)$ and the measurement outcome $\mathbf{s}$. Here we show only the ket layer of the state. The gates are derived from Eq.~\eqref{eqn:ZZgate}~\eqref{eqn:Xgate}. 
    }
    \label{fig:bilayertensornetwork}
\end{figure*}

The 2D surface code can be written as a PEPS~\cite{Gu2009tensornetworkstringnet, Vidal09tensornetworkstringnet, BULTINCK2017183}. 
In this representation, the weak measurement Kraus operators act as random local non-unitary operators on the physical legs, conditioned upon the classical measurement outcomes. For a visual illustration of the resulting PEPS see Fig.~\ref{fig:bilayertensornetwork}.
Generic physical observables can be obtained by taking two copies of the PEPS, as the ket and bra, respectively, and contracting out the physical legs with observable operators inserted.  
The bilayer tensor network, $\braket{\psi(\mathbf{s})}$, then defines the partition function of an associated 2D classical model.
This partition function can be readily obtained from the post-measurement tensor network wavefunction $|\psi \rangle$ in Fig.~\ref{fig:bilayertensornetwork}(a) as the Born probability
\[
	P(\mathbf{s}) = \langle \psi | M^\dagger_{\mathbf{s}} M_{\mathbf{s}} |\psi \rangle = \sum_{\{Z,\tilde{Z}\}} \exp(-E_{AT}) \,,
\] 
where~\cite{teleportcode}
\begin{equation}
\begin{split}
E_{AT,ij} &= - \frac{J s_{ij}}{2} \left(Z_i Z_j + \tilde{Z}_i \tilde{Z}_j\right) + \frac{i \phi}{2} \left(Z_i Z_j - \tilde{Z}_i \tilde{Z}_j\right) \\
&~ - \frac{1}{2} \left( 2K + \frac{i\pi}{2} (1-s_{ij}) \right) \left(Z_i Z_j \tilde{Z}_i \tilde{Z}_j -1\right),
\end{split} \label{eqn:AT}
\end{equation} 
which is a (2+0)D {\sl classical Ashkin-Teller model}~\cite{AshkinTeller43} where $Z_i, \tilde{Z}_i = \pm 1$ are the classical Ising variables of the two layers of the tensor network wavefunction $|\psi (\mathbf{s}) \rangle$ and $\langle \psi(\mathbf{s})|$, respectively.
Here $K, J$ are related to the measurement strength $t$ and coherent rotation angle $\theta$ by $\tanh{(J)} = \sin{(2t)} \cos{(\theta)}$ and $e^{-2K} = \sinh{(J)} \tan{(\theta)}$.
Note that for a finite rotation angle towards the $Y$ direction, $\phi > 0$, this Ashkin-Teller model is augmented by {\sl complex} couplings. In the expressions above, we omit the dependence on the logical qubit and focus on the tensor in the bulk, while the logical qubit can be captured by the boundary virtual leg of the tensor network, see Refs.~\cite{teleportcode,Wang25selfdual} for more details. 

\subsubsection*{(1+1)D quantum circuit \& fermion representation}
The 2D classical tensor network can also be interpreted as a (1+1)D monitored quantum circuit, which evolves towards a boundary quantum chain. 
Its transfer matrix can be written as $T := T_v T_h  = \exp\left(-H_v \right) \cdot \exp \left(-H_h \right)$ where $H_{h,v}$ have been previously derived~\cite{teleportcode} as
\begin{equation}
    \begin{split}
    -H_h &= \sum_j \frac{Js_{2j,2j+1}}{2}(\sigma^x_{2j} \sigma^x_{2j+1}+\sigma^y_{2j} \sigma^y_{2j+1}) \\
    &+i\frac{\phi}{2} (\sigma^x_{2j} \sigma^x_{2j+1}- \sigma^y_{2j} \sigma^y_{2j+1})\\
    &- \left(K+i\frac{\pi}{2}\frac{1-s_{2j,2j+1}}{2}\right)(\sigma^z_{2j} \sigma^z_{2j+1}+1) \ ,
    \end{split}
    \label{eq:Hh_transfer}
\end{equation}
which in turn can be fermionized as
\begin{equation}
    \begin{split}
    &-H_h = \sum_j Js_{2j,2j+1} (c^\dagger_{2j} c_{2j+1} + h.c.) + i \phi (c_{2j} c_{2j+1} + {\rm h.c.}) 
    \\
    &- 2\left(K +i\frac{\pi}{2}\frac{1-s_{2j,2j+1}}{2}\right)(2n_{2j}n_{2j+1}-n_{2j} - n_{2j+1}+1) \, , 
    \end{split}
    \label{eq:Hh_fermion}
\end{equation}
where $n_j = c^\dagger_j c_j$ is the fermion number operator. $H_v$ takes the same form as $H_h$, with the replacement $(2j,2j+1) \to (2j-1,2j)$ and the coupling constants transformed into their Kramers-Wannier dualized versions, c.f.\  Eq.~\eqref{eq:Jdphid} below. Physically, $T_h$ and $T_v$ implement quantum gates on even and odd links, respectively. 

In general, the fermionic Hamiltonian is {\sl interacting}, except when the measurement basis is restricted to $Z$ or $X$ axes. However, we find that a {\sl non-interacting} fermion model can capture the projective-measurement limit for arbitrary measurement angles, see Fig.~\ref{fig:bilayertensornetwork}(b)(c). We postpone the detailed analysis of this scenario to Sec.~\ref{subsec:transfermatrix} below. 

\subsection{Coherent information}

The coherent information, conceptualized in quantum information theory to quantify the size of a coherent logical space through a noisy quantum channel~\cite{Nielsen96coherentinfo}, 
was recently identified as an efficient (scalable) order parameter for measurement-induced phase transitions~\cite{Gullans20scalabledecoder}, and for diagnosing topological quantum memory in the presence of incoherent noise~\cite{Fan23toriccode}, under a teleportation channel~\cite{teleportcode}, or for topological mixed states of fermions~\cite{Huang25coherentinfo}.
Here we employ it as a diagnostic for the topological phase transition of a mixed toric code state under measurement.   
To keep this manuscript self-contained, let us briefly review this concept, explain it in an intuitive manner and -- following its mapping~\cite{teleportcode} 
to the tensor network and statistical mechanical models introduced above, 
which enables us to calculate our phase diagrams to high precision. 

The quantum coherent information is a measure for the quantum information in a system, which can be encoded and decoded. For instance, in the surface code one logical qubit can be encoded and used for further manipulations. After the measurement, the amount of quantum information that remains still inside the code is given by the coherent information~\cite{Nielsen96coherentinfo}. 
To calculate it, we consider three parties of our system as illustrated in Fig.~\ref{fig:fig2projphasediagram}(a):
(i) a reference qubit $R$ that is initially maximally entangled with the logical space of (ii) a surface code $B$ and (iii) an observer $A$  that extracts some information from the measurement outcomes.
Then the coherent information is expressed as the entropy of the reference subtracted by the mutual information between the reference and the observer
\begin{equation}
I_c = S(R) - I(R:A) = S(R A) - S(A) = S(R|A) \ .
\end{equation}
As a conditional quantum entropy, it evaluates how much extra information is needed for the observer to learn the reference qubit. Thus $I_c=0$ means the observer can learn the state of the reference based on the measurement records, while $I_c=\log 2$ means the observer learns nothing about the quantum information based on the measurement outcome. Note that a negative value of $I_c$ cannot appear in our measurement setting (but could in a noisy scenario, where both quantum and classical information are lost~\cite{Nielsen96coherentinfo}).  
In our case the observer has access to a purely classical state only, which can be cast as a diagonal density matrix $\sum_\mathbf{s} P(\mathbf{s})\ketbra{\mathbf{s}}$. 
Further note that the diagnostic depends only on the correlation between $R$ and $A$ (the observer), while the quantum degrees of freedom in $B$ (the surface code) are traced out (discarded):
\begin{equation}
\includegraphics[width=.5\columnwidth]{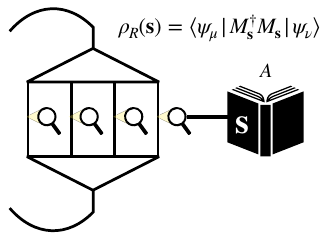} \ .
\end{equation}
Therefore the phase diagram depends essentially only on the ensemble of the reference states
\begin{equation}
\{ P(\mathbf{s}) \ , \rho_R (\mathbf{s}) \} \ ,
\end{equation}
where the state $R$ is obtained by tracing out the measured surface code, whose matrix elements are given by
\begin{equation}
\rho_{R,\mu\nu} (\mathbf{s}) = \bra{\psi_\mu} M^\dagger_\mathbf{s} M_\mathbf{s} \ket{\psi_\nu}\ ,
\label{eq:rho_R}
\end{equation}
where $\psi_{\pm}$ marks the ground state of the surface code with the respective logical qubit state.
Note that this ensemble of states can be viewed as {\it the ensemble of the topological logical qubit} encoded into the surface code.

\begin{figure*}[t] 
    \centering
    \includegraphics[width=\textwidth]{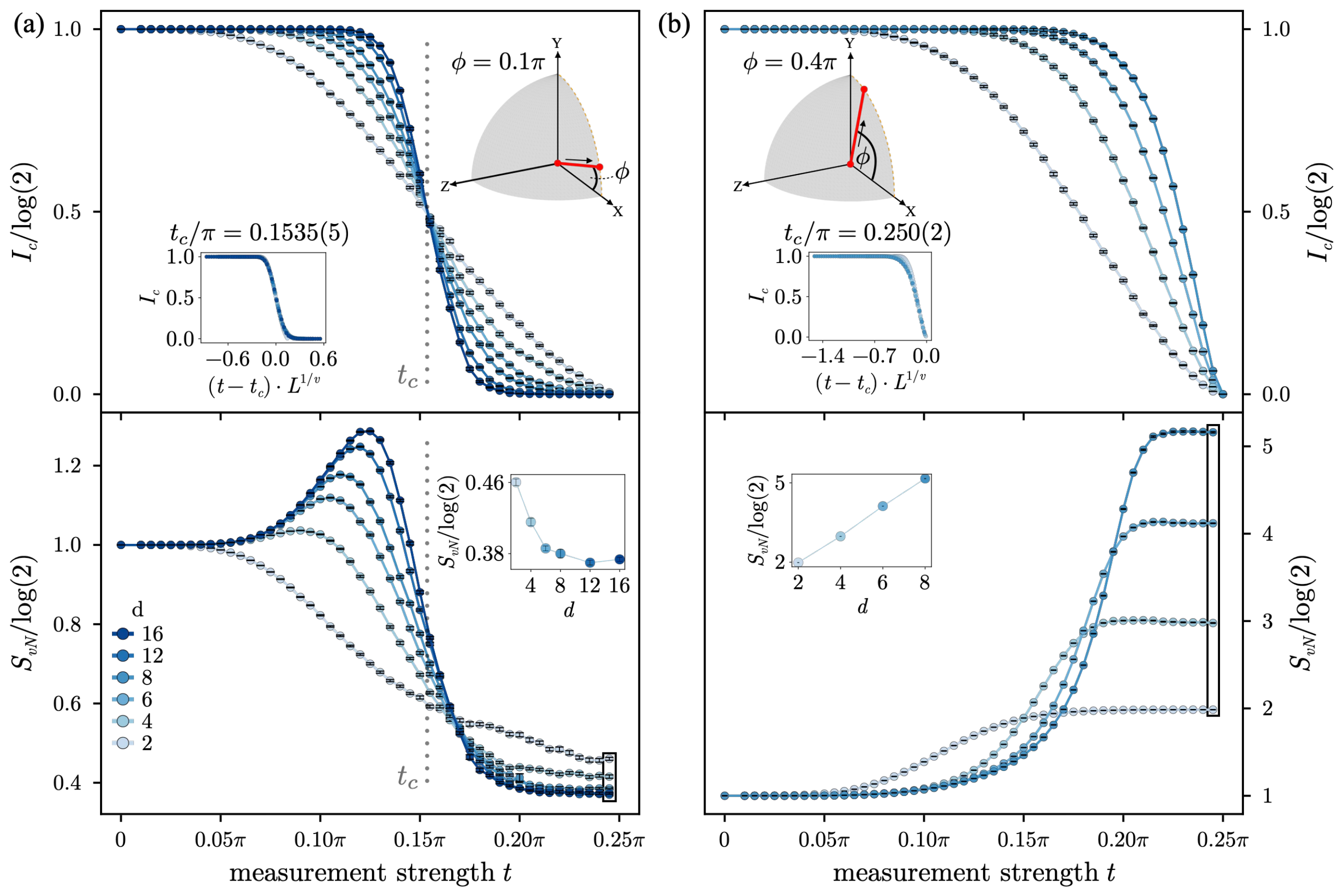}
    \caption{
        {\bf Learning transitions by tuning the measurement strength.}
        Shown are two cuts in the $XY$ plane for {\sl finite} measurement strength $t$, one far from and one close to the $Y$-point, shown on the left/right respectively. 
        The top row shows the {\sl coherent information}, while the bottom row shows the {\sl boundary entanglement entropy.}
        (a) For a cut at an angle $\phi=0.1\pi$ (close to the $X$ point in the $XY$ plane) the transition occurs at a {\sl finite} measurement strength 
        with the coherent information exhibiting a crossing point at a critical $t_c$. 
        The boundary entanglement entropy exhibits a peak that approaches the transition point for larger system sizes with $\log(L)$ scaling,
        characteristic of the scaling at a (1+1)D quantum critical point.
        In the projective-measurement limit of $t=\pi/4$ the entanglement entropy decreases with system size, indicating an area law phase. 
        (b) For a cut at an angle $\phi=0.4\pi$ (close to the $Y$ point in the $XY$ plane) the transition occurs only in the {\sl projective} measurement limit ($t=\pi/4$),
        where the coherent information exhibits a collapse.
        The boundary entanglement entropy, in contrast, exhibits an entanglement {\sl plateau} for finite system sizes, whose height scales linearly with code distance,
        	reminiscent of a volume-law scaling.}        
    \label{fig:WeakMeasurementTransition}
\end{figure*}

We can denote the matrix elements of Eq.~\eqref{eq:rho_R} as
\begin{equation}
    \rho_R(\mathbf{s}) = \frac{1}{2P(\mathbf{s})}\begin{pmatrix}
        P_{++}(\mathbf{s}) & P_{+-}(\mathbf{s}) \\
        P_{-+}(\mathbf{s}) & P_{--}(\mathbf{s})
    \end{pmatrix}
    \equiv \frac{1+\vec{\kappa} \cdot \vec{\sigma}}{2} \,, 
\end{equation}
where the  prefactor is necessary for proper normalization. The probability $P(\mathbf{s})$ of a given set of measurement outcomes can be obtained by $P(\mathbf{s}) = (P_{++}(\mathbf{s})+P_{--}(\mathbf{s}))/2$.
On the right-hand side of the equation above we define an effective polarization field  in the logical space which results form the observer's measurement
\begin{equation}
    \vec{\kappa}(\mathbf{s}) =\left(\frac{P_{++}-P_{--}}{P_{++}+P_{--}} 
\ ,\quad
\frac{2|\Im(P_{+-})|}{P_{++}+P_{--}}\ ,\quad
\frac{2|\Re(P_{+-})|}{P_{++}+P_{--}}
\right) \ ,
\end{equation}
whose strength can be viewed as the effective learning strength in the long-wavelength limit, after tracing out the whole physical degrees of freedom. 
Its angle indicates the {\sl effective measurement angle} of the logical qubit, which should be distinguished from the measurement angle 
of the physical qubits that is fixed when choosing a measurement basis. 
Each component of $\vec{\kappa}$ can be related~\cite{Zhu19} to the anyon condensation and confinement mechanisms at play when decohering the topological quantum order of the toric code: 
The component $\kappa_X$ is a sign of $m$-particle condensation, the component $\kappa_Z$ a sign of $e$-particle condensation and the $\kappa_Y$ component 
can only arise by a mixture of both mechanisms and therefore has to include measurement of $Y$ operators. (Thus $\kappa_Y$ vanishes, for instance, in the $X-Z$ plane.)

Expressed in terms of this field, the coherent information becomes the Born average of the conditional reference entropy $I_c = \sum_\mathbf{s} P(\mathbf{s}) I_\mathbf{s}$, where $I_\mathbf{s}$ is the von Neuman entropy of the reference or logical qubit conditioned upon the observer's measurement outcome $\mathbf{s}$
\begin{equation}
    I_\mathbf{s} = 
- \frac{1 + C_\mathbf{s}}{2} 
\log \frac{1 + C_\mathbf{s} }{2}
- \frac{1 - C_\mathbf{s}}{2}
\log \frac{1 - C_\mathbf{s} }{2} \ ,
\end{equation} 
with $C_\mathbf{s} = |\vec{\kappa}(\mathbf{s})|$ being the strength of the polarization field. 

As it is a measure for the information of the logical qubit, the coherent information is $I_c=\log 2$ in the unperturbed surface code phase and $I_c=0$ in the limit of projective measurement. Where the coherent information drops between these two limiting values, possibly at some intermediate measurement strength, the transition from surface code phase to an ensemble of product states occurs. 
This information-theoretical observable is found to exhibit minimal finite-size effects~\cite{Mueller23coherentinfo} and has provided, for instance, the highest-precision determination of the threshold and critical exponents of the Nishimori phase transition~\cite{Wan25nishimori, Puetz25percolation, teleportcode, Puetz25learning}.
In our interacting fermion case here, this behavior of the coherent information is illustratively shown in the top row of Fig.~\ref{fig:WeakMeasurementTransition}. with panel (a) corresponding to the situation where the transition occurs at {\sl finite} measurement strength, while panel (b) corresponds to a situation where the transition is pushed out all the way to the projective limit.

Recasting the coherent information in the language of a learning transition, the coherent information evaluates the amount of information preserved in the code~\cite{Nielsen96coherentinfo}, that is {\sl not} extracted by the observers. 
In the ``thermodynamic limit" of large code distances ($d\to\infty$), the coherent information converges to a step function that maintains one bit of information ($\log 2$) when the measurement strength remains below the threshold, and zero when it moves above the threshold. 
The precise threshold of the measurement strength can be determined, e.g.\  by performing finite-size scaling (in the code distance), 
and thereby pinpoint the location of the learning phase transition.
We map out such thresholds for arbitrary measurement angles and thereby determine the phase diagram in Fig.~\ref{fig:protocol}(b) above.
For a full spherical illustration see also Fig.~\ref{fig:3DPhaseDiagram} below.
One can immediately recast this phase diagram also in terms of the post-measurement ensemble of quantum states.
Below the measurement threshold, i.e.\ inside the interior of our ``carved sphere", the quantum states remain topologically ordered, 
while the states dephased into classical loop gas~\cite{Wang25selfdual} above the threshold, i.e.\ on the surface of our sphere and in its ``carved-out" regions,
see Fig.~\ref{fig:3DPhaseDiagram}. 
In the language of spontaneous symmetry breaking (SSB), the topological phase exhibits SSB of the underlying 1-form symmetries, 
while the dephased phase exhibits a strong-to-weak SSB of the 1-form symmetries~\cite{Lee23criticalityunderdecoherence,Luo24weaksym, Wang23aversym, Wang24strtowksym, You24weaksym, LiuXuPollmannKnap2025oneform}. 

\subsection{Boundary entanglement entropy}

Our second diagnostic to detect and characterize the critical theory of the learning transition is the boundary entanglement entropy.
The latter is calculated for a one-dimensional chain of boundary qubits that we attach to the surface code and which we leave untouched
when measuring out the physical qubits of the surface code. The entanglement structure of the Choi state 
(obtained by flattening the density matrix to a ``wave function'' vector~\cite{Choi1975}) of this boundary can be interpreted as arising 
from a (1+1)D monitored quantum circuit evolution (induced by the 2D bulk measurements), with the relevant transfer matrices discussed
around Eqs.~\eqref{eq:Hh_transfer} and \eqref{eq:Hh_fermion} above. 
Alternatively, we can think of the boundary entanglement entropy to characterize the computational complexity of contracting out 
the 2D tensor network in Fig.~\ref{fig:bilayertensornetwork}(a).

Numerical results for the boundary entanglement entropy along the two finite-measurement cuts in the $XY$ plane of the phase diagram  
are shown in the lower panel of Fig.~\ref{fig:WeakMeasurementTransition}. 
For the learning transition at finite measurement strength along the first cut (close to the $X$ measurement direction), we observe
a non-monotonic behavior of the entanglement entropy with an intermediate peak close to the transition. For increasing code distances, 
this peak not only moves closer to the transition, but shows a {\sl logarithmic} divergence -- indicative of a (1+1)D quantum critical point. 
In contrast, for the cut close to the $Y$ point where the learning transition is pushed all the way to the projective-measurement limit
the boundary entanglement entropy monotonically increases with no indication of an intermediate transition. Notably, however, an
entanglement plateau is visible whose height increases {\sl linearly} with code distance, reminiscent of a volume-law phase. 
However, since for larger system sizes the width of the plateau also shrinks and possibly vanishes for very large code distances, 
this plateau might not indicate an intermediate phase, but point to interesting behavior in the projective-measurement limit which
we characterize in detail in the following section.

\section{Projective-measurement limit}
\label{sec:projectivelimit}

One of the most striking observations in mapping out the learning transition in the $XZ$ plane has been that at the self-dual $X+Z$ point 
the transition is pushed all the way out to the projective limit \cite{teleportcode, Wang25selfdual}. 
This situation where the surface code remains robust or ``unlearnable" all the way up to the projective-measurement limit no longer remains
a fine-tuned possibility, but instead becomes the generic behavior once we introduce a finite $Y$-measurement component, 
see our phase diagram in Fig.~\ref{fig:protocol} or, probably even more convincing, the 3D illustration in Fig.~\ref{fig:3DPhaseDiagram} below.
But how much do we actually learn about the surface code's logical state in this projective limit? 
For this, it will turn out to be quite instructive to look at the reference qubit which, in the projective-measurement limit, captures the purified logical qubit. 
In particular, we want to understand its distribution when the measurement protocol is repeated for multiple rounds with many copies of the code. 
We will show that ``unlearnable" is a label that only applies to weak measurements, while for projective measurements quite the opposite is true --
by performing multiple measurements along the same measurement angle one can perform {\sl full-state tomography} of the logical qubit state 
encoded in the surface code.

For these generic measurement angles where one is able to learn the logical qubit state via tomography of the reference qubit, 
one might wonder what the nature of the measured-out surface code is in this projective limit. That is, what can we say about 
the ensemble of product states or, equivalently, the set of measurement readout snapshots in the projective-measurement limit? 
Is this ensemble describing a critical state akin to what has been observed for the self-dual $X+Z$ point \cite{Wang25selfdual} 
or is it a general ``gapless" state (with long-range entanglement similar to the ground state of a gapless Hamiltonian)? 
We will show that, generically, one finds the latter situation, i.e.\ a metal-like gapless Gaussian fermion state. 

\subsection{Ensemble of logical snapshots}

\begin{figure}[t!]
    \centering
    \includegraphics[width=.89\columnwidth]{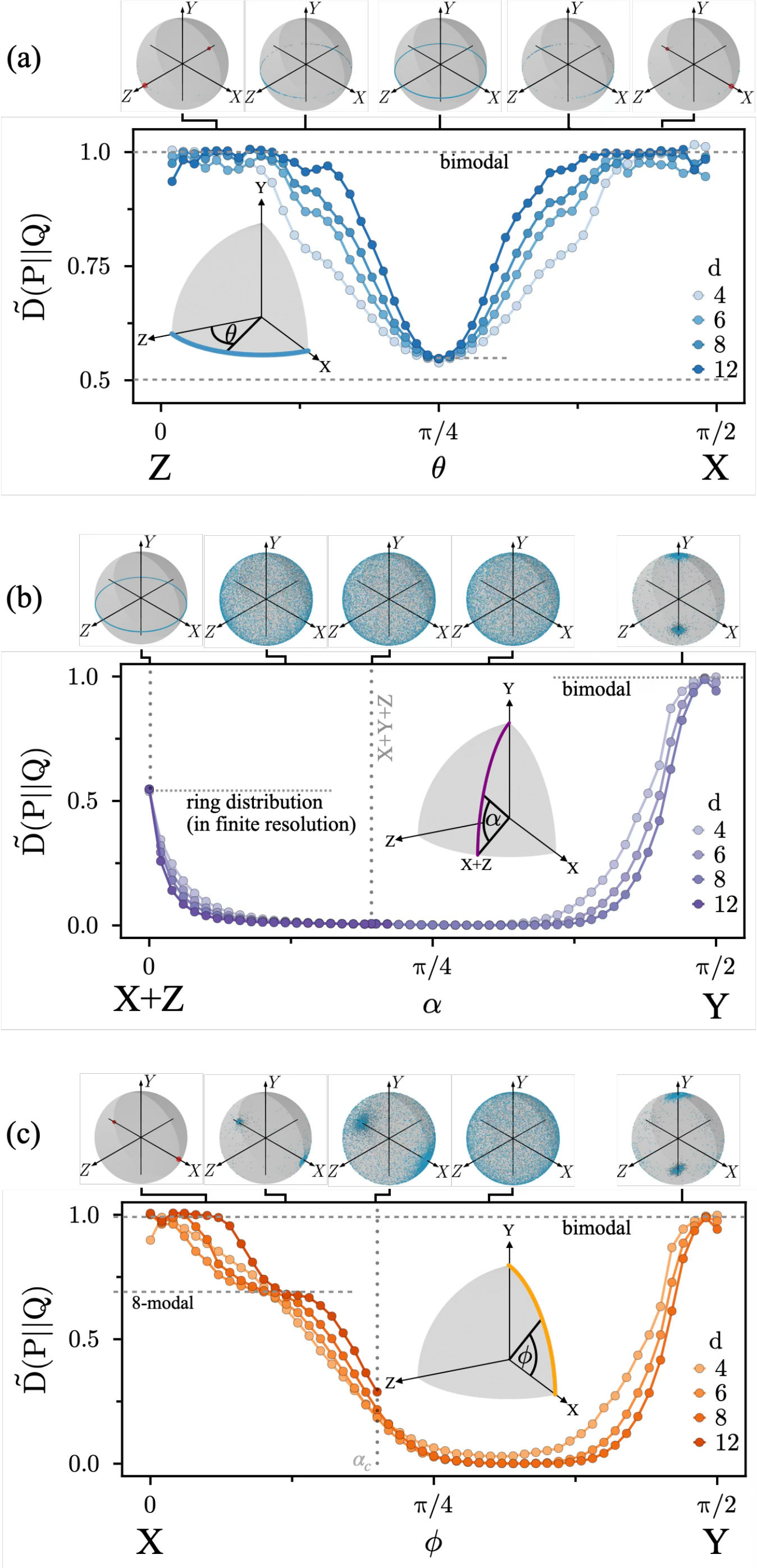}
    \caption{{\bf Projected ensemble of the reference qubit on  (a) the XZ line, (b) the self-dual line, and (c) the XY line.} The spheres above each plot show the projected ensemble distribution for different lines on the Bloch sphere labeled by $\alpha$ in the phase diagram by displaying $10^4$ sampled points from the distribution. The $X$ and $Z$ points on the Bloch sphere of highly concentrated densities are marked with small red spheres.
    In the above plots, the Kullback-Leibler divergence (relative entropy) between the projected ensemble and a uniform (Haar) distribution of the Bloch sphere is normalized by the bimodal distribution value $\tilde{D}(P||Q) = D(P||Q) (\log n - \log 2)$ where $n$ is the number of discrete patches covering the sphere. 
    For the ring distribution, $\tilde{D}(P||Q)$ becomes $1 - \log(m) / \log(n)\to 1/2$, where $m$ is the number of patches covering the equator as a 1D line, such that the ratio of dimension is $\log m / \log n \to 1/2$ for the limit of high resolutions. In the $n \to \infty$ limit, the dip in (a) at $\theta = \pi/4$ and the first data point at $X+Z$ in (b) reach exactly $\tilde{D}(P||Q) = 1/2$. See Appendix~\ref{app:KL_divergence} for more details. 
     }
    \label{fig:KW_divergence}
\end{figure}

Let us start our discussion of the projective-measurement limit, by first noting that the coherent information vanishes in this limit, $I_c=0$, because the surface code state collapses to a classical snapshot $\ket{\mathbf{s}}$. At the same time, the reference qubit $R$, which we introduced to calculate the coherent information, is purified
into a state $\ket{\psi_R(\mathbf{s})}$ where
\begin{equation}
    \begin{split}
&\ket{\psi_R(\mathbf{s})} = \sum_{\mu=\pm} \bra{\mathbf{s}} U(\theta,\phi) \ket{\psi_\mu} \ket{\mu} \ ,\\
\end{split}
\label{eq:psiR}
\end{equation}
with $U(\theta,\phi)$ being the coherent rotation~\eqref{eq:Utheta}. 
The associated density matrix $\ketbra{\psi_R(\mathbf{s})}$ 
\begin{equation}
    \begin{split}
&\ketbra{\psi_R(\mathbf{s})} = \frac{1+\vec{\kappa}(\mathbf{s}) \cdot \vec{\sigma}}{2} 
\end{split}
\end{equation}
can be visualized as a vector which points at the $\vec{\kappa}(\mathbf{s})$ direction in the Bloch sphere. 
In particular, for measurement in the $Z$ or $X$ basis directions, the purified reference qubit $R$ points exactly at the $\pm Z$ and $\pm X$ directions, where the sign depends on the measurement outcome. Rotating the measurement axis to the self-dual $X+Z$ direction, the purified qubit $R$ is restricted to lie in the $XZ$ plane (since the imaginary part has to vanish). 
Plotting our numerical results in Fig.~\ref{fig:KW_divergence}(a), we see that one observes a uniform ``ring-distribution" in the $XZ$ plane.
Moving the measurement direction out of the $XZ$ plane we observe what looks like a uniform spherical distribution for the regimes where the learnability threshold is moved all the way to the projective limit, see Figs.~\ref{fig:KW_divergence}(b) and (c).

To quantitatively understand this behavior, let us formally write the distribution of the reference qubit as
\begin{equation}
    P(\theta,\phi) = \sum_{\mathbf{s}} P(\mathbf{s}) P(\vec{\kappa}|\mathbf{s}) \ ,
\end{equation}
where $P(\vec{\kappa}|\mathbf{s})$ is a conditional delta-distribution since each measurement outcome $\mathbf{s}$ maps to a deterministic $\mathbf{s} \mapsto \vec{\kappa}(\mathbf{s}) = \cos(\theta)\vec{z}+\sin(\theta)\cos(\phi)\vec{x}+\sin(\theta)\sin(\phi)\vec{y}$. Later we will use $\theta,\phi$ to label $\vec{\kappa}$ in  spherical coordinates. 
This distribution we now want to compare with an ideal reference distribution such as a uniform coverage of the Bloch sphere, 
i.e.\ a  Haar-measure random distribution of the single logical qubit, which can be parametrized by 
\begin{equation}
    Q(\theta, \phi) = \frac{1}{4\pi} \sin\theta d\theta d\phi \ .
\end{equation}
The ``distance" between this reference distribution $Q(\theta,\phi)$ and our observed ensemble $P(\theta,\phi)$ can be quantified by their relative entropy, also known as Kullback-Leibler (KL) divergence 
\begin{equation}
    D(P||Q) = \int P(\theta, \phi) \log \frac{P(\theta, \phi)}{Q(\theta, \phi)} d\theta d\phi \ ,
\end{equation}
where $Q(\theta, \phi)$ is the ideal and $P(\theta, \phi)$ is the measured distribution. 
If two distributions perfectly align, their relative entropy vanishes and the KL divergence is zero.

To evaluate the KL divergence numerically, we perform a Born average over the ``free-energy'' difference between the measured distribution and the ideal distribution
\begin{equation}
D(P||Q) = \mathbb{E}_{P}( \log P - \log Q ) \,, 
\end{equation} 
with more details on our numerical procedure provided in Appendix~\ref{app:KL_divergence}.
To discuss our numerical data in Fig.~\ref{fig:KW_divergence}, let us first mention asymptotic results for some limiting cases.   
For the $Z$ and $X$ measurement limits which converge towards bimodal distributions, the KL divergence (with regard to the uniform distribution) diverges as $\log n - \log 2$ where $n$ is the number of discretized patches of the sphere (see Appendix~\ref{app:KL_divergence}). In contrast, near the self-dual $Y$ limit, it is split into 8 hotspots near the $\pm Y$ poles, which are required by the self-dual symmetry that relates $X$ to $Z$. Consequently, the KL divergence near $Y$ limit reduces to $\log{n} - 3 \log{2}$, which still diverges but is always smaller by 2 bits than the $X$ and $Z$ limits, due to the symmetry. 
For the ``critical'' state at the self-dual $X+Z$ measurement angle, the ideal distribution is a uniformly distributed equator-like ring in the $XZ$ plane
\[
    Q(\theta, \phi) =\frac{1}{2\pi} [\delta(\phi)+\delta(\phi-\pi)]  d\theta d\phi \, ,
\]
where $\delta(\phi)$ is the Dirac delta function that integrates to $1$. As shown in Fig.~\ref{fig:KW_divergence}(a)(b), it approaches a half of the value for the uniform spherical distribution. 
The basin plateau of the relative entropy in Fig.~\ref{fig:KW_divergence}(b)(c), where the KL divergence vanishes, corresponds to the uniform spherical distribution.
At these measurement angles, the reference qubit indeed forms a Haar-random post-measurement ensemble. 

Let us now explain what such a Haar-random reference qubit distribution allows us to infer about the original logical qubit. 
In brief, the Haar-randomness of the reference distribution implies that we can effectively measure the logical qubit in {\it random} directions~\cite{Zoller23randomized}, 
even though the physical measurement basis is {\sl uniformly fixed} at a certain angle. The protocol is as follows:
Instead of starting from a maximally mixed logical state,  
consider that one prepares a surface code in some pure logical state, which remains unknown to the observer and that he wants to learn about.
To this end, the observer performs measurements for all physical qubits along an arbitrary (but uniform) angle, e.g.\ along the generic $X+Y+Z$ direction, and thereby obtains a measurement snapshot in the form of a bit string $\mathbf{s}=++--\ldots$. For each possible $\mathbf{s}$, one can invert the isometry encoder (Fig.~\ref{fig:fig2projphasediagram}(a)) to ``decode''~\footnote{This decoder is the inverse of the encoder that maps a logical qubit to a surface code state, so as to map a physical bitstring to a logical snapshot. It should not be confused as the usual decoder in quantum error correction that learns the logical error from syndrome} the corresponding snapshot of the logical qubit $\ket{\psi_L(\mathbf{s})}$, which exactly equals the wave function $\ket{\psi_R(\mathbf{s})}$ in Eq.~\eqref{eq:psiR} given the same $\mathbf{s}$. If we denote the isometry encoder as $U_{\rm enc}$ that maps a single qubit $\ket{\mu}$ to a surface code in the logical $\mu$ state: $\ket{\psi_\mu} = U_{\rm enc} \ket{\mu}$, then the decoded snapshot of the logical qubit follows Eq.~\eqref{eq:psiR} and Fig.~\ref{fig:fig2projphasediagram}(a) to be 
\begin{equation}
\ket{\psi_L(\mathbf{s})}=U_{\rm enc}^\dag U(\theta,\phi)^\dag \ket{\mathbf{s}} \ .
\label{eq:psiL}
\end{equation}
After this decoding step, one thereby gains a {\sl snapshot of the logical qubit}. Multiple rounds of measurements over many copies of the code then effectively  allow measuring the logical qubit in arbitrary direction, forming a classical shadow~\cite{HuangKuengPreskill2020shadowtomography} that is {\sl tomographically complete}~\cite{Zoller23randomized} in this case. That is, one can perform a full state tomography for an arbitrary prepared logical qubit, based on the measurement snapshots at a single measurement angle (uniformly applied to all physical qubits). 
Such uniform basis measurement is experimentally quite feasible in a number of quantum platforms, particularly for large-scale neutral Rydberg atom arrays~\cite{Lukin2023logical}. 
%

\subsection{Monitored (1+1)D quantum circuit representation}
\label{subsec:transfermatrix}

For each measurement record $\mathbf{s}$, the reference state~\eqref{eq:psiR} or the equivalent logical snapshot~\eqref{eq:psiL} can be computed by contracting the tensor network in Fig.~\ref{fig:bilayertensornetwork}, with extra legs that capture the reference qubit (see Ref.~\cite{teleportcode} for details), while the physical qubits are traced out. 
In the projective-measurement limit of interest here, the bilayer tensor network reduces to two {\sl decoupled} layers, as illustrated in Fig.~\ref{fig:bilayertensornetwork}(b). Each layer corresponds to a random classical 2D tensor network state, which can be viewed as a sampled snapshot of the 2D surface code state with coherent error, and the tensor network captures the wave function amplitude of such a snapshot.
Remarkably, each layer of the 2D tensor network can further be mapped to a (1+1)D Ising-type quantum circuit, which can then be mapped to a {\sl free-fermion} circuit. 
Contracting this decoupled tensor network or its free-fermion representation not only gives us direct access to the probability distribution of the measurement outcomes $P(\mathbf{s})$, but also the boundary entanglement entropy of the (1+1)D quantum state. It can also be done in a highly effectively manner~\cite{Wang25selfdual}, allowing us to simulate considerably larger  code distances up to $d=256$ in the projective-measurement limit. 

Let us derive the (1+1)D monitored quantum circuit representation, by identifying step-by-step the quantum gates of this  circuit from the tensor network in Fig.~\ref{fig:bilayertensornetwork}(b). 
Specifically, an elementary gate in this circuit is generated by measuring a qubit on a given bond $(ij)$, where the virtual index of the tensor network is labeled by $\sigma_i$ and $\sigma_j$ for $i,j$ on the adjacent vertices. Under the duality picture, the Pauli $Z$ of the bond qubit is related to the virtual index by $Z_{ij} = \sigma_i \sigma_j$. Physically, $\sigma_i$ and $Z_{ij}$ can be interpreted as the order and disorder (domain-wall) parameters, respectively. Therefore, projective measurement of the bond qubit yields the following weight
\begin{equation} 
\bra{s_{ij}=\pm} e^{i\frac{\theta}{2}Y} e^{i\frac{\phi}{2}Z}\ket{\sigma_i\sigma_j} \propto 
\exp \left(\frac{J s_{ij} + i(\phi-\pi \frac{1-s_{ij}}{2})}{2}\sigma_i \sigma_j\right) \,,
\end{equation}
where the effective measurement strength
\begin{equation}
J(\theta) = \tanh^{-1}\cos\theta
\label{eq:Jtheta}
\end{equation}
ranges from $\infty$ to $0$ as $\theta$ goes from $0$ to $\pi/2$ i.e.\ upon varying the measurement angle from $Z$ to the $XY$ plane. 

Cast in the (1+1)D setting, one can directly write down the 2-body interacting gate (Fig.~\ref{fig:bilayertensornetwork}(c))
\begin{equation}
M_Z = \exp\left(
    \frac{J s_{ij} + i(\phi-\pi \frac{1-s_{ij}}{2})}{2}Z_j Z_{j+1}
\right)\ , \label{eqn:ZZgate}
\end{equation}
which is a complex time evolution generated by Ising interaction and can be decomposed into a unitary part controlled by $\phi$ and a non-unitary (weak measurement) part controlled by $\theta$, followed by a conditional $\pi$-pulse unitary as a local feedback. 
The negative measurement outcome effectively changes the sign of the real coupling constant and adds a $\pi$ pulse to the imaginary coupling constant, which is consistent with the spatial inversion operation for the physical spin eigenstate pointing towards spherical angle $(\theta,\phi)$ : $\theta \to \pi - \theta$, $\phi \to \phi - \pi$.

The transverse-field-generated evolution gate $M_X$ can be derived by viewing $\sigma_i$ and $\sigma_j$ as the states at two adjacent time slices\begin{equation}
    \begin{split}
M_X &\propto 1 +  e^{-(J_d s + i\phi_d)} sX \\
&\propto
\exp\left(
    \frac{J_d s + i(\phi_d-\pi \frac{1-s}{2})}{2} X
\right)
\ ,
    \end{split} \label{eqn:Xgate}
\end{equation}
 which is essentially a {\sl Kramers-Wannier dual} version of the $M_Z$ gate where
\begin{equation}
        \frac{J_d +i\phi_d}{2}= {\rm KW}\left(\frac{J + i\phi}{2}\right) = -\frac{1}{2}\log\tanh\frac{J + i\phi}{2}
        \label{eq:Jdphid}
\end{equation} 
is the Kramers-Wannier (KW) duality relation. 
More explicitly, 
\begin{equation}
    \begin{split}
        J_d &= -\log{\frac{\sqrt{1+e^{4J} - 2e^{2J}\cos{(2\phi)}}}{1+e^{2J}+2e^{J}\cos{\phi}}} \,, \\
        \phi_d &= -\tan^{-1}{\frac{2e^{J}\sin{\phi}}{1-e^{2J}}} \,.
        \end{split}
        \label{eq:KWdual}
    \end{equation}
%

\subsubsection*{Kramers-Wannier duality}

Let us make some general comments on the Kramers-Wannier (KW) duality in our circuit and its manifestation in the phase diagram. %
Firstly, note some useful properties of the KW map for the coupling constants
\begin{equation}
    {\rm KW}(x^*) = {\rm KW}(x)^* \ , \ 
    {\rm KW}(-x) = {\rm KW}(x) - i\pi/2 \ ,
    \label{eqn:complexKW}
\end{equation}
which means KW($\cdot$) commutes with complex conjugation, but flipping the sign is dualized to a phase shift by $\pi/2$.
Therefore, taking the complex conjugate flips the sign of $\phi$ and results in the sign change of $\phi_d$. 
For the $XZ$ plane, $\phi = 0$, and as such, the self-duality condition ${\rm KW}(x)=x$ is {\it real}, with the unique solution $x=\log\sqrt{1+\sqrt{2}}$ corresponding to the $X+Z$ point. Once we turn on the $Y$-component, the electric-magnetic self-duality condition becomes {\it complex} ${\rm KW}(x)=x^*$, which leads to an {\sl extended self-dual line} from the $X+Z$ point to the $Y$ point, along which the measurement angle is symmetric between $X$ and $Z$ 
\begin{equation}
    {\rm KW}(x)=x^* \quad \Leftrightarrow \quad \cos\phi = \cot\theta = \sinh J \ .
    \label{eq:selfdualcondition}
\end{equation}
As is evident from the quantum gates $M_Z$ and $M_X$, in the projective-measurement limit, the (1+1)D quantum circuit can be modeled as a monitored Ising chain, which can be equivalently described by a monitored {\it non-interacting} fermion chain, thereby simplifying the analytical analysis substantially. We will utilize this facilitation to study the non-Hermitian dynamics of the monitored fermion chain under post-selection in Sec.~\ref{sec:postselection}.

The 2D tensor network state and its corresponding (1+1)D monitored quantum circuit share the same probability distribution function
\begin{equation}
    P(\mathbf{s}) = \braket{\psi (\mathbf{s})} = \norm{M_X M_Z \cdots M_X M_Z \ket{\phi}}^2 \, ,
\end{equation}
where $\ket{\psi}$ denote the 2D quantum state (neglecting the logical qubit dependence for simplicity here), while $\ket{\phi}$ denotes the 1D quantum state evolving from the left to the right boundary by the random transfer matrix $M_X$ and $M_Z$ {\sl conditioned} upon the measurement outcomes $\mathbf{s}$. 
Such ``monitored trajectories" that follow Born's probability rule can be efficiently sampled~\cite{Wang25selfdual} 
and are in one-to-one correspondence with the measured bit strings $\mathbf{s}$ of the 2D quantum state.

Before we proceed to a presentation of our numerical results for this, let us look at the structure of the quantum gates $M_X$ and $M_Z$ 
of the (1+1)D monitored circuit for three special limits: 
(i) the self-dual line from the self-dual $X+Z$ to the $Y$ point,  
(ii) the $Y$ point, and 
(iii) the $XY$ line from the $X$ to the $Y$ point.

\begin{figure}[tb!]
    \centering
    \includegraphics[width=\columnwidth]{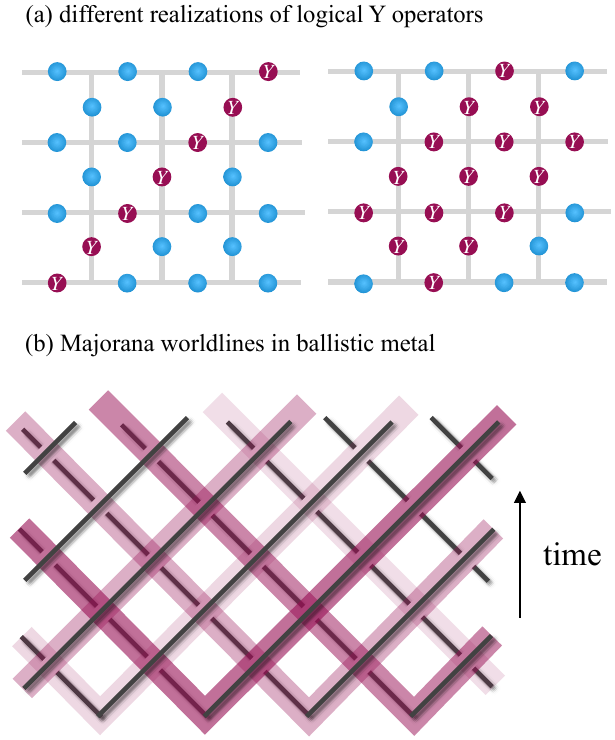}
    \caption{{\bf Projective Y-measurement limit.}
        (a)~Logical Y operators. By measuring the Pauli $Y$ operators on the purple qubits, one can learn the logical $Y$ information, i.e.\ $Y_L = \pm 1$.  
        (b)~Ballistic metal from measuring toric codes in the $Y$ basis. The light cones in different shades illustrate the effective (1+1)D dynamics of Majorana. The measurement operators in Eq.~(\ref{eqn:YpointMajorana}) become purely unitary and act as swap operators between maximally entangled Majorana pairs. As a result, the entangled Majorana pairs labeled by the same color propagate ballistically in time along the light cones. 
    }
    \label{fig:ballisticmetal}
\end{figure}

\subsubsection*{Self-dual line}
Along the self-dual line from the $X+Z$ to the $Y$ point, parametrized by Eq.~(\ref{eq:selfdualcondition}), the (1+1)D quantum circuit exhibits Kramers-Wannier self-duality.
It generalizes the measurement-only circuit at the $X+Z$ point, extensively studied in Ref.~\cite{Wang25selfdual}, to one which, in addition, includes unitary gates. 
This can be seen by plugging the self-dual condition into Eq.~\eqref{eq:KWdual}, such that $J_d = J$ and $\phi_d = -\phi$, and the quantum gates of the (1+1)D quantum circuit turn into
\begin{equation}
    \begin{split}
    M_Z &= \exp\left(
        \frac{J s_{ij} + i(\phi-\pi \frac{1-s_{ij}}{2})}{2}Z_j Z_{j+1}
    \right)\ ,\\
    M_X &=
    \exp\left(
        \frac{J s - i(\phi-\pi \frac{1-s}{2})}{2} X
    \right) \ ,
    \end{split}
\end{equation}
which are invariant under translation and sublattice complex conjugation. 
Whereas at the $X+Z$ point, where $\phi = 0$ the circuit reduces to a measurement-only quantum circuit,
once we turn on the $Y$ component along the self-dual line, the circuit becomes a hybrid one of measurements and unitary evolutions.

\subsubsection*{Exactly solvable dual unitary dynamics in the Y-limit} 
\label{subsubsec:Y-point}
In the $Y$-only measurement limit, $J=0=J_d$ and $\phi=-\phi_d=\pi/2$, the gates reduce to purely unitary gates of $\pi/2$ pulses (maximally entangling gates) with random signs
\begin{equation}
    M_Z \to e^{\pm i \frac{\pi}{4} Z_jZ_{j+1}} \ , \quad M_X \to e^{\pm i \frac{\pi}{4} X} \ .
    \label{eqn:YpointMajorana}
\end{equation}
This circuit dynamics is shown in Majorana representation in Fig.~\ref{fig:ballisticmetal}(b) where the ``light cone" dynamics illustrates the formation of a {\sl ballistic metal}. It is a special case of the dual-unitary circuit, which is unitary along both the ``time" direction and the ``space" direction. Consequently, the entanglement entropy grows as
\begin{equation}
    S_{vN} = t \log 2 \ ,
\end{equation} for $d>t$.  

In this limit, all the physical qubits are collapsed onto the $Y$ basis. Without extra noise, one can read the logical $Y$ information from the measurement outcomes supported on the logical $Y$ operators. See Fig.~\ref{fig:ballisticmetal}(a).

\begin{figure*}[t!]
    \centering
    \includegraphics[width=2\columnwidth]{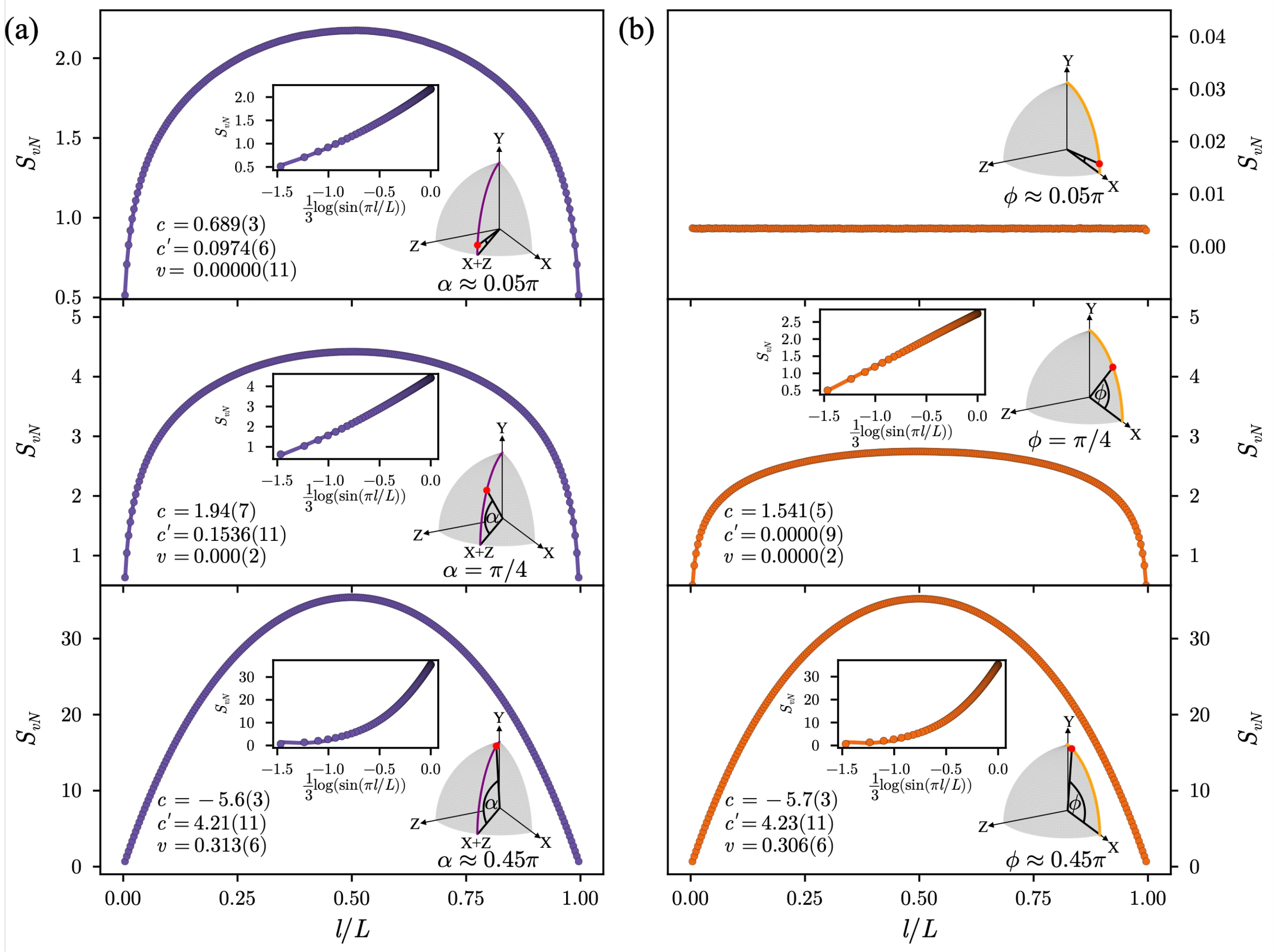}
    \caption{{\bf Entanglement arcs}. 
    Shown are numerical data for the boundary entanglement entropy $S(l)$ as a function of subsystem size $l$ along the boundary of total length $L=256$ 
    at different points in the projective-measurement limit.
    The solid lines are the fitting curves using the entanglement arc formula of Eq.~(\ref{eq:entanglementarc_formula}). 
    The prefactors obtained by these fits are given and the insets show the data rescaled by $1/3 \log(\sin(\pi l/L))$. 
    The entanglement arcs are shown for measurement angles on (a)~the self-dual line and (b)~the $XY$ line, 
    where their measurement angle values are indicated by red dots in the inset. 
    }
    \label{fig:entanglementarcs}
\end{figure*}

\subsubsection*{XY line}
\label{sec:XYline}

Along the $XY$ line, $\theta=\pi/2$, which leads to $J=0$ and $J_d=-\log\tan\frac{\phi}{2}$ and $\phi_d=-\pi/2$. Consequently, the quantum gates reduce to
\begin{equation}
    \begin{split}
    M_Z &= \exp\left(
        i\frac{\phi}{2}Z_j Z_{j+1}\right)
    \ ,\\
    M_X &=
    \exp\left(\frac{s J_\phi}{2} X\right) \ ,
    \end{split}
\end{equation}
where $J_\phi\equiv -\log(i\tan\frac{\phi}{2})=\tanh^{-1} \cos\phi - i \frac{\pi}{2}$. 
Here all the random $Z_j$ operators inserted into the circuit are propagated to the final time slice, by flipping certain signs of the $X$ evolution. 
The corresponding 2D statistical model can then be simply written as 
\begin{equation}
\mathcal{Z}(\mathbf{s}) \propto \sum_\sigma \exp \left( \frac{i}{2}\sum_{ij} (\phi -\pi \frac{1-s_{ij}}{2}) \sigma_i \sigma_j \right) \ ,
\end{equation}
which can also be easily derived from the tensor network [c.f.~Fig.~\ref{fig:bilayertensornetwork}]. For example, the Bolztmann weight for a single bond can be derived as follows
\begin{equation}
\includegraphics[width=.8\columnwidth]{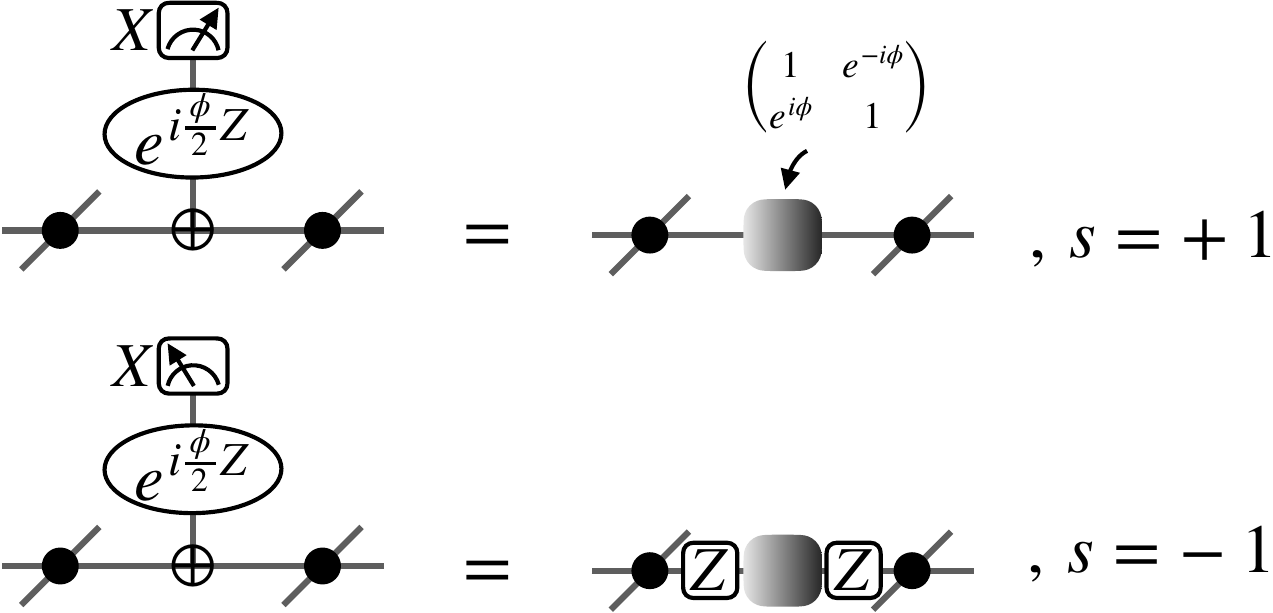} \ .
\end{equation}
Let us close this detailed description of our circuit model, with the observation that the exact same tensor network and statistical model also describes the phase transition in a model where the coherent rotation is followed by a subsequent {\sl syndrome} measurement~\cite{bravyi2018correcting, Beri23toriccodecoherenterror, Beri24coherenterror}. 
We will further discuss such connections with other models in our general discussion section at the end of the manuscript.

\begin{figure*}[th!] 
    \centering
    \includegraphics[width=.98\textwidth]{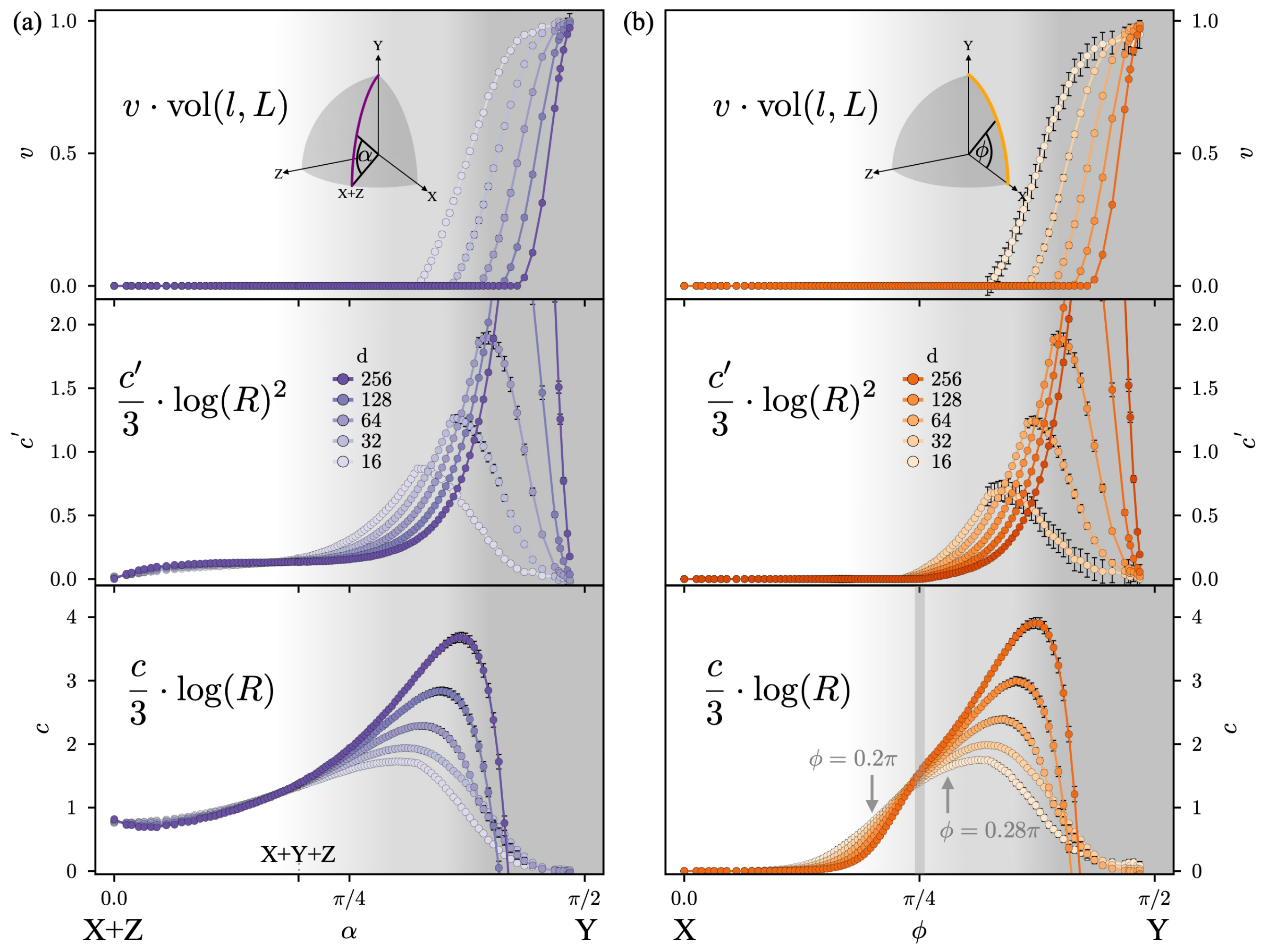}
    \caption{{\bf Boundary entanglement entropy for the projective-measurement limit on (a)~the self-dual line and (b)~the XY line. }
    The prefactors are the result of fitting the numerical data for Eq.~(\ref{eq:entanglementarc_formula}). (a)~On the self-dual line, the system always shows the critical behavior $c \cdot \text{log} (R)$ in the entanglement entropy before reaching the volume-law entanglement regime in a finite system size. In addition, we also find a non-zero contribution from $c' \cdot \text{log}(R)^2$ once we turn on Y-component for $\alpha >0$. 
    Both of these contributions are stable for small $\alpha$ and diverge before vanishing in the volume-law entanglement phase. 
    There is a clear volume-law entanglement phase around $\alpha=\pi/2$. However, this regime shrinks with increasing system size. In the thermodynamic limit, the volume-law behavior only appears at the Y-point. The specialty of the Y-point is discussed in detail in Sec.~\ref{subsubsec:Y-point}.
    For the $XY$ line, the crossing point of $c$ is taken as an indication of a metal-insulator transition around $\phi_c \approx 0.25\pi$. 
    For comparison, we also indicate the critical value $\phi=0.28\pi$ obtained by conductivity calculations in Ref.~\cite{Beri23toriccodecoherenterror},
     and the phase boundary of the infinite-threshold phase $\phi_c \approx 0.2 \pi$ from the small-size interacting-fermion coherent information calculations 
     in Fig.~\ref{fig:protocol}(b).
     }
    \label{fig:coefficients_cut_selfdual}
 \end{figure*}

\subsection{Boundary entanglement entropy}
\label{subsec:entanglemententropy}
Let us now investigate the boundary entanglement entropy of the (1+1)D quantum circuit and map out the phase diagram shown in 
Fig.~\ref{fig:fig2projphasediagram}(c).
Numerical results for the boundary entanglement entropy in the projective-measurement limit along (a)~the self-dual line and (b)~the XY line are shown in Figs.~\ref{fig:entanglementarcs} and \ref{fig:coefficients_cut_selfdual}, respectively. 
To identify different contributions to their entanglement structure we fit the entanglement entropy to a formula
distinguishing volume-law contributions for $v \neq 0$, from metallic  $c' \neq 0$ and critical $c\neq 0$ contributions
\begin{equation}
    \label{eq:entanglementarc_formula}
    S_{vN}(l,L) = v \cdot \textrm{vol}(l,L) + \frac{c'}{6} \cdot (\log{R})^2 + \frac{c}{6} \cdot \log{R}  + a \,,
 \end{equation}
where $R=\frac{L}{\pi}\sin{(\pi \cdot \frac{l}{L})}$ is the chord length of a length-$l$ segment on a circle of circumference $L$, 
and $\textrm{vol}(l,L)$ is the Page curve of Gaussian fermions~\cite{Kieburg21pagefermion}. 
As discussed in Sec.~\ref{sec:summary}, the vol$(\ell,L)$ contribution stems from the ballistic spreading of entanglement under unitary evolutions. 
The $(\log R)^2$ term is known to arise in the Clifford-dynamics of monitored Majorana, which is a setting that can be mapped to the physics of completely packed loop models with crossings (CPLC)~\cite{Nahum13cplc, Fisher21measurementcriticality, Hsieh23toriccode, KlockeBuchhold2023majorana}. 
This contribution is also found to be indispensable in our {\sl non-Clifford} scenario of Gaussian fermions (which cannot be captured by a statistical loop model).
The remaining $\log{R}$ term in Eq.~(\ref{eq:entanglementarc_formula}) is the standard universal logarithmic behavior associated with an underlying conformal criticality of the quantum circuit~\cite{CalabreseCardy2004entanglement,CalabreseCardy2009cftentanglement}. 

As Fig.~\ref{fig:entanglementarcs} and \ref{fig:coefficients_cut_selfdual} show, 
the onset of $Y$ component generally increases the entanglement entropy. For the self-dual line, small deviations from $X+Z$ critical point (which exhibits logarithmic  entanglement scaling only) quickly turn on squared logarithmic scaling as the leading term. This readily indicates that the whole self-dual line is immersed in the gapless phase (Fig.~\ref{fig:fig2projphasediagram}(c) purple region). For the $XY$ line, on the other hand, it takes a finite deviation $\phi>0$ to leave the gapped phase transitioning into the gapless phase -- where we use ``gapped'' and ``gapless'' to refer to short-range entangled and long-range entangled phases, respectively. 
Locating the precise angle where this entanglement transition occurs, can be done in two complementary ways:
From the entanglement-entropy scaling in the projective-measurement limit, we obtain a (rough) estimate of the transition point $\phi_c \approx 0.25\pi$, where it displays an almost perfect logarithmic scaling (as shown in the middle panel of Fig.~\ref{fig:entanglementarcs}(b)). 
This point also coincides with a crossing point in the finite-size scaling fit of $c$ (for code distances up to $d=256$), 
see the bottom panel in Fig.~\ref{fig:coefficients_cut_selfdual}(b) above. 
In comparison, our earlier estimate of the boundary of the phase with infinite (measurement-strength) threshold was $\phi_c \approx 0.2 \pi$ (see Fig.~\ref{fig:WeakMeasurementTransition}(b)), 
where we looked at codes up to distance $d=16$. The same transition point has been estimated from conductivity calculations in Ref.~\cite{Beri23toriccodecoherenterror} to be around $\phi_c \approx 0.28\pi$. 
The discrepancy here is likely due to the strong finite-size effects of the underlying metal-insulator-transition, which requires further investigation.
Put together, the existing estimates indicate that the transition point should lie somewhere within the window $0.20\pi \lesssim \phi_c \lesssim 0.28\pi$ as we have indicated by the two arrows in the lower panel of Fig.~\ref{fig:coefficients_cut_selfdual}(b).
Coming close to the $Y$ point, there appears to be an intermediate volume-law scaling regime, although the volume-law prefactor vanishes in the thermodynamic limit everywhere except the exact $Y$ point. This is attributed to the formation of a ballistic 2D metal in the $Y$-limit as discussed above.

Stepping back and looking at our overall phase diagram, let us make the following general observations:
The most striking feature of our phase diagram is an extended phase with long-range entanglement. Its origin can be best understood as a metallic phase 
that arises in the following way: From the original tensor network representation of our learning problem, we have derived a (1+1)D monitored free-fermion circuit
(introduced in Sec.~\ref{subsec:transfermatrix} above). Its non-unitary dynamics in (1+1)D can be mapped to the Hermitian ground state physics of a Born-disordered fermion system in (2+1)D spacetime
~\cite{Jian2023criticality, Wang25selfdual, Ludwig25volume}. 
In this mapping, the majority of our phase diagram, the phase with long-range entanglement charted out above, corresponds to a {\sl Majorana metal} phase (generally in symmetry class DIII in the Altland-Zirnbauer classification)~\cite{Jian2022criticality, Nahum23majorana, Jian2023criticality, Fidkowski23majorana}, described by the $SO(N\to 1)$ nonlinear $\sigma$ model. 
In contrast, the gapped phase with short-range entanglement corresponds to an insulator, which undergoes a topological transition~\cite{ReadGreen2000} across the self-dual $X+Z$ point, which separates the insulating phases in the $X$ and $Z$ limits, respectively~\cite{Wang25selfdual}. Note that the $XZ$ line~\cite{Wang25selfdual}, $XY$ line~\cite{Beri23toriccodecoherenterror} and $ZY$ line are special limits that fall into symmetry class D with broken time reversal symmetry (for each layer).

One should also draw a distinction of our phase diagram from the Clifford phase diagrams~\cite{Hsieh2021measure, Fisher21measurementcriticality} that is mapped to a classical statistical loop model~\cite{Nahum13cplc}. In particular, the Clifford phase diagram describes a distinct problem to ours, where the bulk qubits of the surface code are subjected to random Clifford projective measurements in the $X, Y, Z$ directions~\cite{Hsieh2021measure}. 
As a result, their resulting phase diagram and the universality class of the phase transitions are distinctive from ours. 
%

\section{Non-Hermitian dynamics from post-selection}
\label{sec:postselection}
\begin{figure*}[t!]
    \centering
    \includegraphics[width=2\columnwidth]{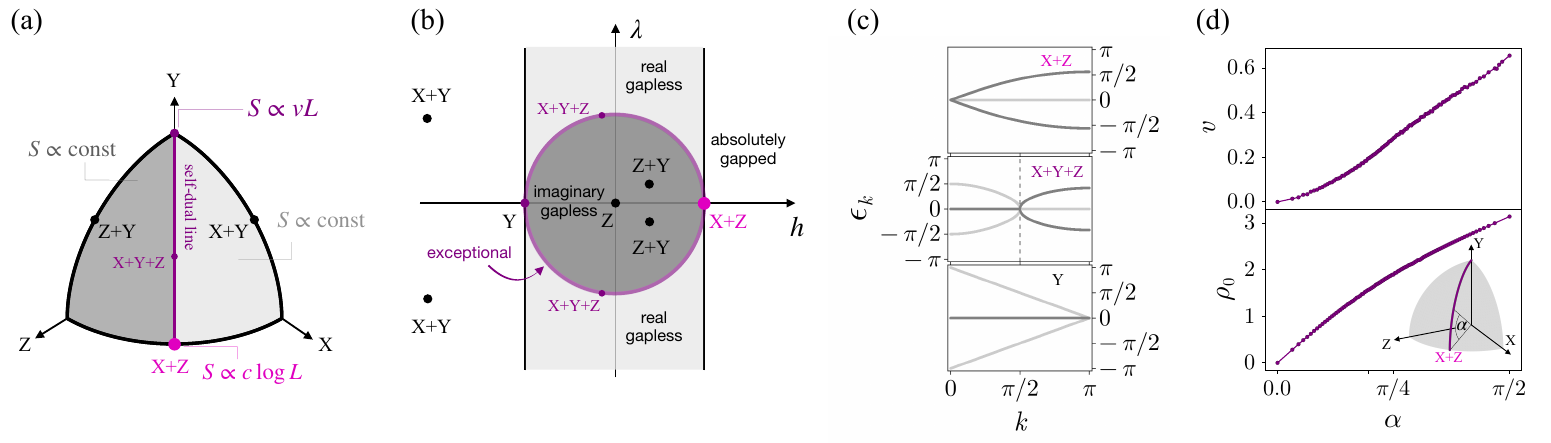}
    \caption{{\bf Non-Hermitian physics from post-selection}. 
        (a) Phase diagram characterized by the boundary entanglement entropy for post-selection limit 
        [c.f.~Fig.~\ref{fig:fig2projphasediagram}~(b) in comparison with Born-average limit]. 
        The purple color, for finite $Y$-component, indicates a {\it volume-law} behavior in boundary entanglement entropy which persists even in the thermodynamic limit, which is absent in the Born scenario. 
        For the post-selection phase diagram, dark and light gray regimes are long-range and short-range entangled area-law phases in the qubit representation, respectively. 
        In the Majorana fermion representation, they correspond to topological and trivial Majorana chains, respectively. 
        (b) Phase diagram of the effective non-Hermitian Majorana chain Hamiltonian Eq.~(\ref{eqn:Heff}). 
        In particular, the self-dual line in (a) maps to the unit circle of the exceptional points in (b).  
        The dark and light gray regions of the post-selection diagram in (a) map to the inside and outside the unit circle in the non-Hermitian Majorana phase diagram, respectively. 
        (c) Spectrum of the Floquet operator $U_F = \prod_k e^{-i H_k}$ along the self-dual line, 
        where the Majorana fermion is described by the non-Hermitian Hamiltonian $H_k$, with $k$ denoting the momentum. 
        The dark and light gray lines represent the real and imaginary parts of the spectrum $\epsilon_k$ of $H_k$. 
        (d) The volume-law prefactor $v$ from boundary entanglement entropy [Eq.~(\ref{eq:entanglementarc_formula})] 
        and the density of steady states $\rho_0$ from the spectrum of Floquet operator [Eq.~(\ref{eqn:FloquetU})] along the self-dual line are correlated. 
        The boundary entanglement entropy was calculated with system size $L= 128$ and the spectrum of the (1+1)D free fermion Floquet operator was calculated with system size $L = 10,000$. 
     }
    \label{fig:postselect}
\end{figure*}
{\sl Post-selection} can provide a powerful complement to the Born-averaged description of quantum dynamics with measurement that we have discussed so far. 
While Born averaging yields an unconditional, trace-preserving evolution that captures what happens in reality, it washes out structures that exist at the level of individual trajectories. In contrast, conditioning on measurement outcomes (i.e.\ performing post-selection), especially in the projective-measurement limit, turns the same measurement process into a selective probe that isolates specific sectors of the trajectory ensemble, making otherwise invisible dynamical patterns transparent.
As a result, post-selection yields a deterministic, ``clean'' trajectory with greater analytic tractability than the Born-averaged evolution. 
In this sense, post-selection can act as a microscope to resolve the mircostructure of monitored dynamics beyond typical (Born-averaged) runs.

\subsection{Non-unitary Floquet operator}
Building on the projective-measurement limit discussed in Sec.~\ref{sec:projectivelimit}, we investigate the post-selection limit,
where the dynamics are conditioned on a fixed sequence of measurement outcomes rather than Born-averaging over all possible measurement outcomes $\{\mathbf{s}\}$ in our qubit measurements of the surface code.
To be specific, we focus on the post-selected configuration where $\mathbf{s} = +1$ for all measurement outcomes in the following discussion.
In this case, the effective (1+1)D quantum circuit in Fig.~\ref{fig:bilayertensornetwork}(c) is simplified to a Floquet operator
\begin{equation}
    U_F = \exp\left(
        \frac{J  + i\phi}{2}\sum_j Z_j Z_{j+1}
    \right)
    \exp\left(
        \frac{J_d + i\phi_d}{2}\sum_j X_j
    \right)
    \ , \label{eqn:FloquetU}
\end{equation}
which realizes a non-Hermitian Floquet Majorana / Ising model~\cite{Ravindranath2023, Su2024}, c.f.~Eq.~(\ref{eqn:ZZgate}) and (\ref{eqn:Xgate}). In the Majorana representation, the Floquet operator becomes
\begin{equation}
    U_F = \exp \left( \frac{J+i\phi}{2} \sum_j i \gamma_{2j} \gamma_{2j+1} \right) \exp\left( \frac{J_d + i\phi_d}{2} \sum_j i \gamma_{2j-1} \gamma_{2j} \right) \,.
    \label{eqn:FloquetUMajorana}
\end{equation}
The phase diagram of the non-Hertimian Flouqet Ising chain characterized by the boundary entanglement entropy of the steady state is shown in Fig.~\ref{fig:postselect}(a).
Notice that when $|J  + i\phi| = |J_d  + i\phi_d|$, the model is (complex) Kramers-Wannier self-dual in the sense of Eq.~(\ref{eqn:complexKW}) and can be fermionized as a Majorana chain with complex hopping constants and with no dimerization. 
In the post-selection limit, volume-law scaling only appears on the self-dual line except at the $X+Z$ point, at which the system is critical with log behavior in the entanglement entropy. Away from the self-dual line, the dark and light gray regimes represent long-range and short-range
entangled area-law phases, respectively. By contrast, in the Born-averaged case, we have seen that finite code distances lead to a small, but finite regime of volume-law behavior in the entanglement entropy (which disappears only in the ``thermodynamic limit" of large code distances). 
Notably, the post-selection and Born-average pictures are naturally 
connected in {\sl replica field theory}. Specifically, the post-selection limit and Born-average correspond to the $n \to \infty$ and $n \to 1$ limits, respectively, where $n$ is the replica number. As $n$ increases, the volume-law regime shrinks and eventually turns into a single (self-dual) line as $n \to \infty$.

\subsection{Effective non-Hermitian Hamiltonian and its phase diagram}

Besides the boundary entanglement entropy,
the spectrum of the Floquet operator $U_F$ also reveals rich structures. 
In particular, the {\it effective Hamiltonian} of the Floquet operator $U_F$ in Eq.~(\ref{eqn:FloquetUMajorana}) in the limit $|J+i\phi|,~|J_d + i\phi_d| \ll 1$ can be obtained via the Baker-Hausdorff formula as
\begin{equation}
    \begin{split}
        H_{\text{eff}} &= - \sum_j i \gamma_{2j} \gamma_{2j+1} - (h+i\lambda) \sum_{j } i\gamma_{2j-1} \gamma_{2j} \,,
    \end{split} \label{eqn:Heff}
\end{equation}
where $h + i \lambda = \frac{J_d + i\phi_d}{J + i \phi}$. 
This limit can be realized, e.g.\ by choosing distinct measurement angles for the qubits on the horizontal and vertical bonds of the surface code, by recalling $J(\theta) = \tanh^{-1}\cos\theta$ from Eq.~\eqref{eq:Jtheta} and Eq.~\eqref{eq:Jdphid}.
The phase diagram of $H_{\text{eff}}$ is shown in Fig.~\ref{fig:postselect}(b). Specifically, the dark gray region inside the unit circle is gapless in the imaginary part but gapped in the real part of the spectrum, while the light gray region outside the unit circle is gapless in the real part but gapped in the imaginary part. By contrast, for large magnetic field, $h>1$, the spectrum is absolutely gapped, i.e.\ gapped in both real and imaginary parts.
In the effective Hamiltonian $H_{\text{eff}}$, $|h+i\lambda|=1$ is the {\it complex} self-duality condition, while $h=1$ is the {\it real} self-duality condition. [See also the discussion below Eq.~(\ref{eqn:complexKW})].
Particularly, the self-dual line maps to the unit circle of {\it exceptional} critical points of $H_{\text{eff}}$; the long-range and short-range area-law phases in Fig.~\ref{fig:postselect}(a) map to the inside and outside the unit circle, respectively.
Notice that the lower half of $(h,\lambda)$-plane is redundant, as $\lambda \to -\lambda$ is equivalent to $(J,\phi) \to -(J,\phi)$, which does not change the phase.  In particular, at the self-dual $X+Z$ point, $J=J_d$, $\phi = \phi_d = 0$, corresponding to the $(1,0)$ point in the $(h,\lambda)$ plane; at the $Y$ point, $J=J_d = 0$, $\phi= -\phi_d = \pi/2$, corresponding to the $(0,1)$ point; at the $X+Y+Z$ point, $J=J_d = \tanh^{-1}(1/\sqrt{3})$ and $\phi = -\phi_d = \pi/4$, which is also labeled on the unit circle in Fig.~\ref{fig:postselect}(b).  
Other special points off the self-dual line in Fig.~\ref{fig:postselect}(a) can be readily determined as well. Specifically, the $X$ point is mapped to infinity on the $(h,\lambda)$ plane; 
the $Z$ point is mapped to the origin; the $X+Y$ point is mapped to the absolutely gapped region and the $Z+Y$ point is mapped inside the self-dual unit circle. 

\subsection{Quasi-energy spectrum and its connection to boundary entanglement entropy}
The Floquet operator for the non-interacting Majorana fermions can be readily diagonalized in momentum space as $U_F = \prod_k e^{-i H_k}$, where $H_k$ is the Floquet Hamiltonian with quasi-energies $\epsilon_k$. (A detailed analysis of the Floquet operator is relegated to Appendix~\ref{app:nonhermitianfermion}.) 
Notably, zero modes of Im$(\epsilon_k)$ correspond to the steady states that contribute to entanglement spreading. 
At the self-dual $X+Z$ point, the (1+1)D Floquet Majorana state exhibits a logarithmic entanglement entropy scaling with the code distance $d$, which is related to the fact that the Floquet spectrum has a steady state only at $k = 0$ as shown in Fig.~\ref{fig:postselect}(c); other imaginary modes correspond to the forced-measurement-only circuit that evolves the quantum state to the low-energy area-law corner of the Hilbert space. 
By contrast, the onset of a $Y$ component ($\phi, \phi_d \neq 0$) turns the imaginary time into a complex-time evolution. 
As such, there exist states whose quasi-energies are purely real, [c.f.\ Fig.~\ref{fig:postselect}(d) at the $X+Y+Z$ and $Y$ points].  As discussed before, these are the steady states that contribute to the spreading of entanglement entropy. Consequently, the late-time-evolved state has volume-law entanglement entropy behavior.
At the $Y$ point,  Im$(\epsilon_k)$ always vanish. As such, the system has the largest density of steady states contributing the entanglement spreading. Notice that the real energy gap closes at the $k=\pi$ momentum point. At this momentum point, we have Majorana zero modes $(\epsilon_{k = \pi} = 0)$; $U_F$ is reduced to the exchange operators of these Majorana fermions, as discussed below Eq.~(\ref{eqn:YpointMajorana}).  

To understand further the connection between the steady states from the spectrum and the volume-law behavior in the boundary entanglement entropy, 
we show the volume law scaling prefactor $v$ and the density of steady states defined as $\rho_0 = \frac{1}{\pi} \sum_k \delta(\text{Im}(\epsilon_k))$ along the self-dual line in Fig.~\ref{fig:postselect}(d). 
Notably, the steady state density is nonzero {\it only} along the self-dual line, which agrees with the nonzero volume-law factor $v$ obtained from
 the boundary entanglement entropy [c.f.~Fig.~\ref{fig:postselect}(a)]. Notice that the increase of $\rho_0$ also qualitatively agrees with that of $v$. Intuitively, the steady states of the non-Hermitian dynamics do not decay under the dissipative dynamics. Consequently, their ballistic motion contributes to the entanglement spreading described by quasiparticle picture~\cite{calabresecardy2005quench}. 
 The quantitative deviation comes from the fact that the steady states do not contribute equal amount of entanglement entropy~\cite{LuGrover2021spacetimeduality}.   

\subsection{Non-Hermitian interacting fermion chain}
For the finite-strength weak measurement, the post-selected effective Hamiltonian is a non-Hermitian but interacting fermion system. 
The partition function after post-selecting $s_{ij}=1$ can be readily obtained from Eq.~(\ref{eqn:AT}) as 
\[
	P(\mathbf{s}=1) = \langle \psi| M^\dagger_{\mathbf{s}=1} M_{\mathbf{s}=1} |\psi\rangle = \sum_{Z,\tilde{Z}} \exp(-E_{AT}(\mathbf{s}=1)) \,, 
\]
where
\begin{equation}
\begin{split}
E_{AT,ij}(s_{ij} =1 ) &= - \frac{J }{2} (Z_i Z_j + \tilde{Z}_i \tilde{Z}_j) + \frac{i \phi}{2} (Z_i Z_j - \tilde{Z}_i \tilde{Z}_j) \\
&~ - K (Z_i Z_j \tilde{Z}_i \tilde{Z}_j -1) \,.
\end{split}
\end{equation} 
Under the quantum-classical statistical mechanics mapping, the (2+0)D classical AT model becomes a (1+1)D quantum $XXZ$ chain. Specifically,
\begin{equation}
        P(\mathbf{s} = 1) \propto \exp(-\sum_j H_{2j,2j+1}) \cdot \exp(-\sum_j H_{2j-1,2j}) \,,
\end{equation}
where~\cite{teleportcode} 
\begin{equation}
    \begin{split}
    H_{2j,2j+1} &= - \frac{J}{2} (\sigma_{2j}^x \sigma^x_{2j+1} + \sigma^y_{2j} \sigma^y_{2j+1}) \\
    &~ + i \frac{\phi}{2} (\sigma_{2j}^x \sigma^x_{2j+1} - \sigma^y_{2j} \sigma^y_{2j+1}) - K(\sigma_{2j}^z \sigma^z_{2j+1} + 1) \,, \\
    H_{2j-1,2j} &= - \frac{J_d}{2} (\sigma_{2j-1}^x \sigma^x_{2j} + \sigma^y_{2j-1} \sigma^y_{2j}) \\
    &~ + i \frac{\phi_d}{2} (\sigma_{2j-1}^x \sigma^x_{2j} - \sigma^y_{2j-1} \sigma^y_{2j}) - K_d(\sigma_{2j-1}^z \sigma^z_{2j} + 1) \,,
    \end{split} \label{eqn:XXZchain}
\end{equation}
where $\mathbf{\sigma}$ are the spin-1/2 operators of the effective quantum $XXZ$ chain.
Here $K, J$ are related to the measurement strength $t$ and coherent rotation angle $\theta$ by 
$\tanh{(J)} = \sin{(2t)} \cos{(\theta)}$ and $e^{-2K} = \sinh{(J)} \tan{(\theta)}$, and 
related to $K_d, J_d$ via Eq.~(\ref{eq:KWdual}).
The system has a U(1) symmetry when either $J$ or $\phi$ is zero; on the other hand, when both $J$ and $\phi$ are nonzero, the U(1) symmetry of the $XXZ$ chain is broken to a $Z_2$ symmetry.
This $U(1)$ to $\mathbb{Z}_2$ symmetry breaking can be more transparently seen from the fermion representation of the $XXZ$ chain.
Under a Jordan-Wigner transformation, the $XXZ$ chain can be mapped to a complex fermion chain $c_j$ with non-Hermitian pairing terms and density-density interactions. 
Specifically,  Eq.~(\ref{eqn:XXZchain}) becomes
\begin{equation}
    \begin{split}
        H_{2j,2j+1} &= J (c^\dagger_{2j} c_{2j+1} + \text{h.c.}) + i \phi (c_{2j} c_{2j+1} + \text{h.c.}) \\
        &~ - 2K (2n_{2j} n_{2j+1} - n_{2j} - n_{2j+1} +1), \\
        H_{2j-1,2j} &= J_d (c^\dagger_{2j-1} c_{2j} + \text{h.c.}) + i \phi_d (c_{2j-1} c_{2j} + \text{h.c.}) \\
        &~ - 2K_d (2n_{2j-1} n_{2j} - n_{2j-1} - n_{2j} +1),
    \end{split} \label{eqn:fermionXXZ}
\end{equation}
where $n_j = c^\dagger_j c_j$ is the fermion number operator. 
As is evident from the spin-1/2 representation in Eq.~(\ref{eqn:XXZchain}), the U(1) symmetry (corresponding to the fermion number conservation) is broken to $\mathbb{Z}_2$ symmetry (corresponding to the fermion number {\it parity} conservation) when both $J$ and $\phi$ are nonzero. 
Notice that when $J=0$ and $\phi \neq 0$, the system still has a U(1) symmetry after a relabeling of particles and holes on a sublattice, e.g.~$c_{2j} \to c^\dagger_{2j}$.
\\

Let us conclude this section with some comments on the post-selection limit. Post-selected trajectories can qualitatively capture the Born-averaged behavior when the chosen trajectory is typical, i.e., it carries a probability weight that does not decay exponentially with system size or time. 
However, its entanglement dynamics can drastically differ from that of the Born-averaged case if the post-selection trajectory is atypical.
Nevertheless, as we have shown here, the post-selection limit provides a useful lens to understanding the entanglement dynamics of monitored quantum circuits via the non-Hermitian Hamiltonian framework.

\section{Discussion}

Let us close our manuscript, by putting our results into a broader context. 
We start by discussing the connection of our learning phase diagram to studies of noisy channels, the ensuing decoding problems,
and their computational complexity. 
We also draw a connection to other models discussed in the literature. This includes models studying the effect of syndrome measurements 
in the presence of coherent rotations, which we recast to learning problems and establish a one-to-one correspondence to our model for a 
limited set of parameters. We also comment on connections of our model to explorations of non-Hermitian, interacting fermion chains.
We then briefly touch on how our learning phase digram can be put to work in engineering thresholds for teleportation protocols that have
to deal with imperfectly created Bell pairs. 
Finally, we compare the generic measurement-induced deformation of the toric code and its phase diagram studied in this manuscript 
to previously studied deformations, particularly Hamiltonian and wavefunction deformations, and describes similarities and distinctions
in their respective phase diagram.
We end with a hypothesis of how our results might generalize for arbitrary CSS codes.

\subsection*{Monitored dynamics, noisy channel, and decoding}

If one discards the measurement record, e.g.\ by summing over the measurement outcomes, then our model turns into a non-Clifford channel $\bigotimes_j \mathcal{N}_j$ where each qubit is subjected to a dephasing in the $\sigma_j^{\theta,\phi}$ basis
\begin{equation}
\mathcal{N}_j(\cdot) = \cos^2(t) (\cdot) + \sin^2(t) \sigma_j^{\theta,\phi} (\cdot)\sigma_j^{\theta,\phi} \, .
\label{eq:noise}
\end{equation}
In the projective-measurement limit $t=\pi/4$, the maximally dephased mixed state is purely ``classic" and becomes equivalent to the ensemble of post-measurement states -- the core object of this paper. It implies that the toric code subject to maximal dephasing in a generic basis (far away from X and Z axes) turns into a 2D metal state. 
The phase diagram for finite probability of dephasing, however, is not exactly solved yet and it remains an open question to be discussed in (our) future work.
Relatedly, a 2-replica phase diagram of a noisy toric code where each qubit suffers from an i.i.d. single-qubit unitary rotation was studied in Ref.~\cite{lee2024mixedstatetopologicalordercoherent}. It is important to note that such a 2-replica problem is quite distinct from the Born measurement scenario with its correlated disorder structure and distinctive critical phenomena described by (non-unitary) conformal field theories~\cite{Wang25selfdual}. 

\begin{figure}[t]
    \centering
    \includegraphics[width=\columnwidth]{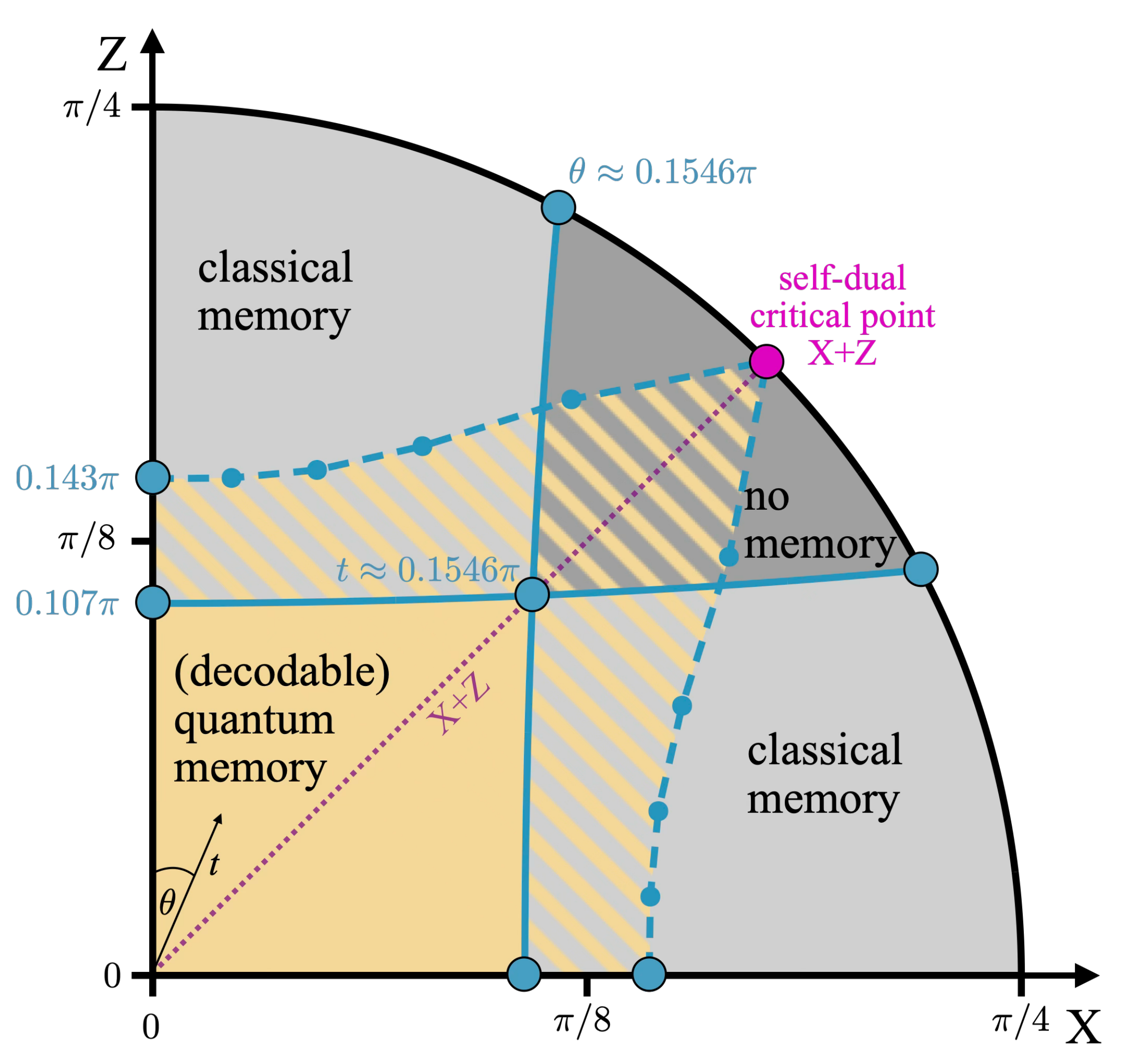}
    \caption{
        {\bf Phase diagram for decodability of the logical information of the surface code.} The dashed blue lines show the transitions for the optimal threshold, and the solid blue lines show the minimal weight perfect matching (MWPM) decoder: 
        In the yellow region MWPM can retrieve the full logical information. In the light grey region, the surface code cannot be used as quantum memory anymore, but still serves as a classical memory using the respective encodings for the X and the Z direction. 
        In the dark grey region, the surface code does not serve as either quantum or classical memory. 
        The regime with yellow stripes between the solid and the dash blue lines mark the regime where the logical information generally is still decodable, but a MWPM decoder fails. 
        All blue lines and points are characterized by Nishimori transitions. The self-dual critical point ($X+Z$) is captured by an unstable RG fixed point in a distinct universality class, which flows away from this unstable fixed point along the Nishimori points under RG flows. 
         }
    \label{fig:decoding}
\end{figure}

For the noisy channel, one can recast the learning problem as a decoding challenge: If the quantum information is still coherently encoded in the dephasing mixed state, can one decode the state and correct to transform it back to a pristine surface code? A straight-forward approach, which however is not guaranteed to produce an optimal result, is to
adopt a minimum-weight-perfect-matching (MWPM) decoder~\cite{Preskill2002}.
Based on a sequence of syndrome measurements it identifies Pauli string operators to pair up the $e$ and $m$ particles separately, so as to correct the surface code after the noise~\eqref{eq:noise} back to the pristine surface code. Since one pairs up $e$ and $m$ separately, the phase diagram for such a decoder generally consists of four phase regions, see Fig.~\ref{fig:decoding} for the XZ plane for example~\cite{Xia25experiment}. The solid blue lines correspond to the Nishimori decoding transition for either $e$ or $m$ string proliferation -- which, however, sits {\sl below} the coherent information threshold (dashed blue line) indicating a suboptimal decoding. 
This suboptimality is particularly striking when turning to the self-dual $X+Z$ direction (dotted purple line) where the threshold is slightly enhanced (compared to the decoding for $X$ or $Z$ noise only), but significantly smaller than the maximal threshold obtainable for an optimal decoder (as indicated by the coherent information threshold at $t=\pi/4$ along the self-dual line). Instead, for sufficiently large $t$, the logical quantum memory is completely lost (not even allowing for a classical memory phase). The reason is that by discarding the entire measurement record (of single-qubit measurements) and treating the $e$ and $m$ particles (syndromes) separately, one cannot capture any of their correlations which become particularly pronounced in this region of the phase diagram.
This points to an interesting open question to develop a more efficient decoder that takes into account both syndrome and qubit measurements and capable enough to detect their underlying correlations.

\subsubsection*{ Decoding complexity}
An alternative approach to the decoding problem would be to employ a tensor network decoder. 
Provided the measurement outcome fed, e.g.\ from a quantum experiment, one can contract out the random tensor network in Fig.~\ref{fig:bilayertensornetwork} to decode its bulk correlations and thereby the observed snapshot of the logical qubit.  
With the analytic insight that the tensor network for projective measurements (i.e.\ in the $t=\pi/4$ limit) can be free-fermionized, one can always contract it with {\sl polynomial} complexity. 
However, for more general scenarios (where one does not have such insight), or if $t<\pi/4$ where the fermions become interacting, the contraction complexity generically depends on the bond dimension of the tensor network representation and thus upon the entanglement entropy of the boundary state~\cite{Harrow2022, NishimoriCat}.  
In particular, in the self-dual plane spanned by the $X+Z$ and $Y$ bases, there are different categories of contraction complexity as summarized in Tab.~\ref{tab:selfdual} below. Overall, staying closer to the $X+Z$ direction implies lower complexity of tensor network contraction, due to the proximity with the ``gapped insulating'' phases in the $X$ and $Z$ neighborhoods. In contrast, while the $Y$ limit is integrable, its entanglement entropy grows and a naive tensor network contraction is exponentially difficult.

\begin{table}[h!]
    \centering
    \begin{tabular}{c | c | c | c}
        \toprule
        angle & $X+Z$ & $X+Z+ \epsilon Y$  & $Y$  \\
        \midrule
        $S_{vN}$ scaling & $\log L$ & $(\log L)^2$ & $L$  \\
        MPS bond dimension $D$ & $L^{c/6}$ & $L^{\log L}$ &  $e^L$ \\
        contraction complexity & polynomial & superpoly & exponential \\
        \bottomrule
    \end{tabular}
    \caption{{\bf Tensor network computational complexity for sampling the toric code in the self-dual directions}.  
    	The top row specifies the measurement direction. 
	In the first row, $S_{vN}$ indicates the boundary entanglement entropy, and $D$ refers to the virtual bond dimension of the boundary MPS when contracting out the tensor network. 
    	While the whole self-dual plane yields infinite threshold, the decoding is not necessarily equally hard. This decoding complexity is indicated in the final row, 
	in terms of the tensor network simulation of the toric code state measured in the given basis, which ranges from polynomially easy in the exact $X+Z$ direction, 
	to superpolynomial when the direction deviates from $X+Z$ with finite Y component. 
    }
    \label{tab:selfdual}
\end{table}

\begin{figure*}[th!]
    \centering
    \includegraphics[width=\textwidth]{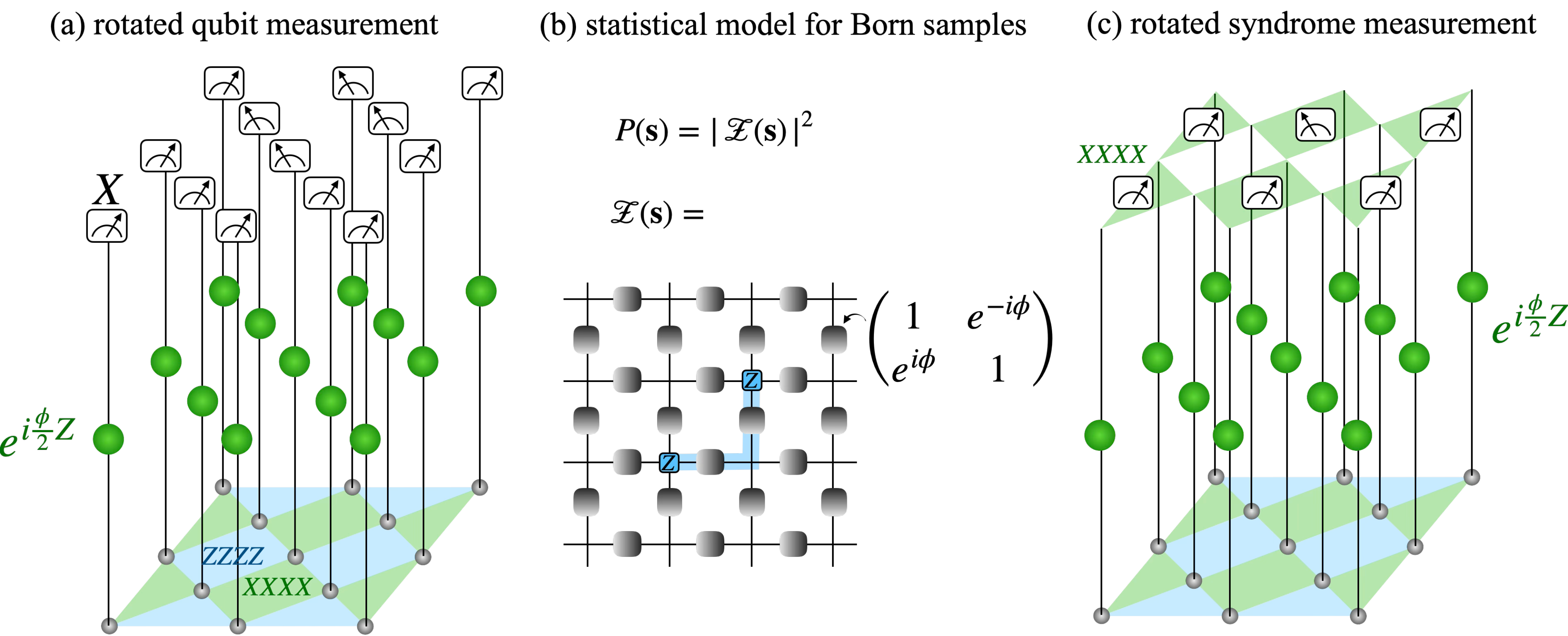}
    \caption{{\bf Qubit versus syndrome measurements.} The same classical statistical model (b) describes the Born sample of measuring either (a) the qubit or (c) the syndrome, of the surface code with a coherent Pauli $Z$ rotation (for all qubits). The tensor network of the statistical model [c.f.~Fig.~\ref{fig:bilayertensornetwork}(a)] is shown in (b), where the random disorder corresponds to injecting Pauli $Z$ operator, or equivalently, a classical Ising variable, to the site where one observes an $e$ particle as $XXXX=-1$ syndrome in (c), or where the string of $X=-1$ measurement outcomes terminates in (a). 
    }
    \label{fig:equivsyndromemeas}
\end{figure*}

\subsection*{Connection to other models}

\subsubsection*{Learning from syndrome measurements}
While we have concentrated on learning the toric code via measurements of its physical qubits, we find that there is a closely related setting
where, instead, one performs standard {\sl syndrome} measurements. At first sight, these situations might look entirely unrelated, 
since syndrome measurements are typically performed with the idea of learning the noise instead of the logical qubit in order to perform quantum error correction, i.e.\ to restore the logical state of the toric code without learning it (so as to preserve it). However, if one performs a coherent rotation on the physical qubits and then performs syndrome measurements, this situation can drastically alter -- with syndrome measurements ultimately destroying the toric code state and instead allowing an observer to learn the logical state via the set of collected syndromes.

Specifically, we have been able to establish a one-to-one correspondence (of the bulk ensemble) between qubit and syndrome measurement for the case of projectively measuring the toric code in the $\cos\phi X+ \sin\phi Y$ single-qubit basis (i.e.\ along the $XY$ line in our phase diagram), 
which leads to exactly the same underlying statistical mechanics model with  that of a toric code under coherent error $e^{i\phi Z / 2}$ followed by a {\sl syndrome} measurement, as illustrated in Fig.~\ref{fig:equivsyndromemeas} above.
This syndrome measurement setting has been studied in Refs.~\cite{bravyi2018correcting, Beri23toriccodecoherenterror, Beri24coherenterror}, while similar setups with general coherent errors are also studied in Refs.~\cite{ Iverson2020, Chen25coherenterror}. 
This not only means they share the same phase transition and criticality, but they also lead to exactly corresponding Born samples, which allow one to learn about the logical information. 
In the previous works~\cite{ Beri23toriccodecoherenterror, Beri24coherenterror}, the phase transition induced by cranking up the rotation angle was interpreted as an error-induced breakdown of the logical qubit. 
However, distinct from the incoherent-noise induced transition~\cite{Preskill2002} where information is completely lost into the environment, in the coherent error scenario the logical information is not all lost, but part of it can be reconstructed by the syndrome observer in the form of a classical shadow. 
This might sound contradictory to the usual notion that the syndrome space is designed to be orthogonal with the logical space, such that it only reveals the noise information without harming the logical space. However, the coherent error rotates the basis and overlaps the syndrome with the logical qubit. One can therefore expect that beyond a rotation angle threshold, the Born samples from the {\it rotated} syndrome measurement allow one to construct a tomographically complete classical shadow for the logical qubit -- a scenario which we will address, in more detail, in future work. 

\subsubsection*{Post-selected non-hermitian interacting fermion chain}
Away from the projective-measurement limit, the post-selected (1+1)D quantum circuit is effectively described by a non-Hermitian 
{\sl interacting} fermion chain, see Eq.~\eqref{eqn:fermionXXZ} above. 
In this regime, the standard quasiparticle picture of entanglement spreading is no longer reliable, since interactions invalidate a simple free-quasiparticle description. Nevertheless, we believe that a robust volume-law phase persists: instead of being confined to the single self-dual line shown in Fig.~\ref{fig:postselect}(a), the region exhibiting volume-law scaling extends over a finite portion of the post-selected phase diagram. 
Figure~\ref{fig:WeakMeasurementTransition}(b) illustrates this behavior on the $XY$ plane, where the Born-averaged entanglement entropy displays clear volume-law scaling away from the projective-measurement limit in finite system sizes. For properly chosen post-selected measurement trajectories -- for example, by conditioning on the most probable measurement configuration -- we therefore expect this volume-law behavior to be captured quantitatively within the effective non-Hermitian interacting fermion chain description.

\subsubsection*{Teleportation transition threshold engineering}
Our global phase diagram in Fig.~\ref{fig:protocol} not only describes the toric code under weak measurement, but also captures the teleportation transition of a toric code with non-maximally entangled resources~\cite{teleportcode, Xia25experiment}. 
In that setting, the parameter $t$ tunes a coherent error in the entangling gate of Bell state preparation for a teleportation protocol
that employs one entangled Bell pair to teleport the physical qubits of the toric code and thereby the full logical qubit state in a robust manner~\cite{teleportcode}. 
Recast in this teleportation protocol, our finding of a substantial infinite-threshold phase region for generic measurement angles, 
dictates that an optimal way to teleport the topological toric (or surface) code is to rotate the relative basis of any imperfectly entangled Bell pairs. Then for generic angle far from the $X$ and $Z$ limits it would suffice to teleport with even {\it arbitrarily weakly entangled pairs}.

\begin{figure}[tb]
    \centering
    \includegraphics[width=\columnwidth]{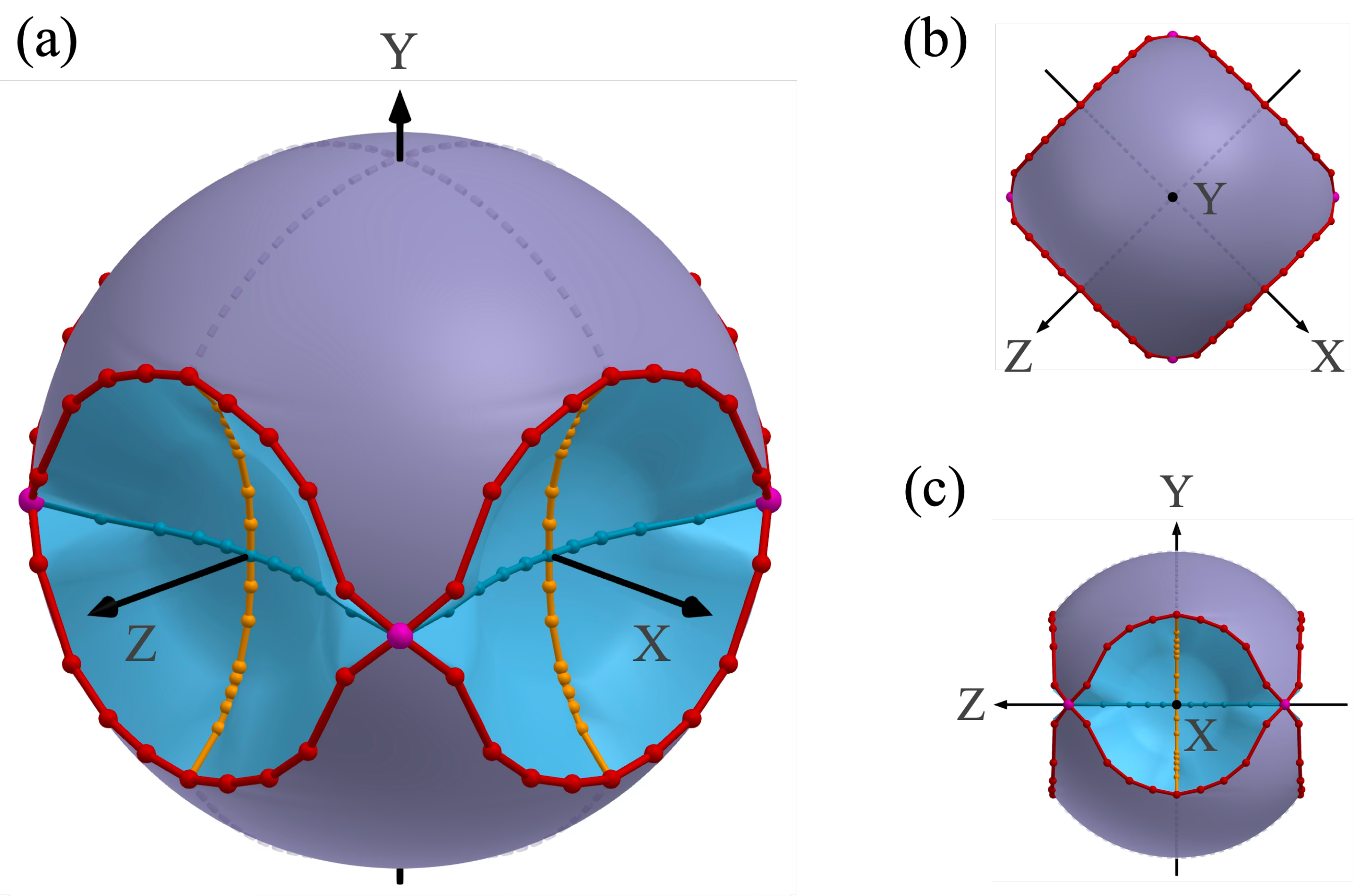}
    \caption{ 
        {\bf Three-dimensional phase diagram} of our model, viewed from different angles. 
        The four blue regions carved out of the sphere correspond to the learnable phases, where the observer successfully learns  
        the logical $Z$ or $X$ information (while collapsing the quantum loop condensate wave function to the corresponding 
        classical loop gas ensemble at some finite measurement strength). 
        For generic directions, the learnability threshold is pushed to the projective-measurement limit (indicated by the purple sphere),
        with the observer obtaining, in the absence of any readout errors, full tomographic information of the logical qubit 
        from repeated measurements, without the need for randomized measurements. 
        From the free-fermion perspective the purple sphere corresponds to a metallic phase in symmetry class DIII 
        -- with the exclusion of the two dashed ``meridian" lines crossing in the $Y$ point indicating symmetry class D, 
        which, in addition, also applies to the ``equator", i.e.\ the projective-measurement limit in the $XZ$ plane.
        }
    \label{fig:3DPhaseDiagram}
\end{figure}

\subsubsection*{Comparison to other toric code deformations}
Let us close by comparing our measurement-induced deformation of the toric code with the deformations previously studied in the literature, i.e.\ the Hamiltonian and wavefunction deformation of the toric code \cite{Troyer10topocrit}, and compare the respective phase diagrams. 
The most general Hamiltonian phase diagram~\cite{Trebst07toriccode, Tupitsyn2010, Dusuel2011}, induced by applying a 3-component magnetic field, is a three-dimensional generalization of the long-known Fradkin-Shenker phase diagram of the Z$_2$ gauge theory~\cite{FradkinShenker79}, giving rise to a box-like phase region~\cite{Dusuel2011} in which the toric code remains robust. The phase transitions at these finite thresholds are either continuous (in the $XZ$ plane) or first-order (in the presence of a $Y$-component). 
By contrast, the wave function deformation phase diagram -- which corresponds to a post-selection scenario from the perspective of measurements -- exhibits not only finite thresholds but also a critical line extending all the way out to infinity (i.e.\ the projective-measurement limit) along the self-dual direction within the $XZ$ plane~\cite{Zhu19}.  Adding a $Y$-component to the measurement basis, this critical point crosses over into a line (in the post-selected, projective-measurement limit) with volume-law entanglement entropy scaling in its 1D boundary state.

The Born measurement phase diagram, which we have explored in this manuscript, is drastically distinct from these previous cases. Instead of breaking down into trivial paramagnet or critical states at high symmetry points, the toric code and its topological quantum order are found to remain maximally stable against non-projective measurements along generic angles. This maximal stability of the topological quantum state has a nice interpretation in the context of the learning setup described in this manuscript, as it results from the maximal input that a  dual classical observer requires in order to learn such a macroscopically entangled quantum state.

Finally, let us mention that there are two natural Born measurement variants of our model and phase diagram: 
(i) Given that qubit loss or leakage error are relatively common in today's NISQ platforms, one might wonder what happens to our model when replacing the measurement {\sl strength} by a measurement {\sl probability}~\cite{Puetz25percolation, Yoshida24monitoredcode}. In such a modified protocol, the Nishimori learnability transition along $X$ or $Z$ measurement axes is replaced by a percolation transition at $50\%$ probability~\cite{Puetz25percolation, Yoshida24monitoredcode}. But for a coherent rotation to a generic measurement direction, when the problem is no longer describable by a Clifford circuit, the ensuing criticality and potential renormalization group flow remain an open question. 
(ii) In further modifying (and simplifying) our model, one could study a {\sl Clifford version} of our spherical phase diagram, 
where also the coherent rotation is replaced by simply unbalancing the projective-measurement probabilities of $X$, $Y$ and $Z$ only. 
While the ``surface" of such a phase diagram, i.e.\ $p(X)+p(Y)+p(Z)=1$, was studied in Ref.~\cite{Hsieh23toriccode}, a global phase diagram similar to our three-dimensional one has not been explored to the best of our knowledge. 
We leave the exploration of these alternative models to future study. 

\subsubsection*{Generalization to CSS code}
Taking a step back and considering our learning phase diagram on the three-dimensional Bloch sphere as illustrated in Fig.~\ref{fig:3DPhaseDiagram},
it becomes strikingly apparent that the learning threshold is {\it generically} infinite. That is, the quantum code is typically stable in an absolute manner, 
because the learning observer (or an adversary eavesdropper) needs full and faithful (projective) measurements in order to read out the complex information 
encoded in the long-range entanglement of the toric code.
The exceptions, however, are the corners surrounding the $X$ and $Z$ axes, where the finite thresholds carved out of the sphere, indicate that an incomplete (weak) measurement already allows for the readout of partial information in the form of the logical $X$ or $Z$ state, while losing the complementary information in the collapse
of the quantum code. This qualitative feature might be generalized to generic CSS codes~\cite{Calderbank1996, Steane1996}, built from $X$ and $Z$ stabilizers. Namely, they could also be stable against generic non-Clifford measurements away from the $X$ and $Z$ corners. We leave this as an interesting subject for future study.

\section*{Acknowledgments}
ST and GYZ would like to thank Andreas Ludwig for illuminating discussions of the metal phase and symmetry class and for collaboration on related projects. 
Two of us (ST and GYZ) are grateful to the Kavli Institute for Theoretical Physics (KITP), which is supported by the National Science Foundation by grant NSF PHY-2309135, and the KITP Programs ``Noise-robust Phases of Quantum Matter" and ``Learning the Fine Structure
of Quantum Dynamics in Programmable Quantum Matter," where part of this work was carried out.
The Cologne research group is supported, in part, by
the Deutsche Forschungsgemeinschaft (DFG, German Research Foundation) under Germany’s Excellence Strategy—Cluster of Excellence Matter
and Light for Quantum Computing (ML4Q) EXC 2004/1 -- 390534769 as well as within the CRC network TR 183 (Project Grant
No. 277101999) as part of subproject B01.
GYZ acknowledges the support of NSFC-Young Scientists Fund (No. 12504181) and Start-up Fund of HKUST(GZ) (No. G0101000221).
Our numerical simulations were performed on the JUWELS cluster at the Forschungszentrum Juelich, 
the Noctua2 cluster at PC2 in Paderborn, and the RAMSES cluster at RRZK Cologne.

\section*{Data availability}
The numerical data shown in the figures and for the ensembles of the reference qubit are available on Zenodo~\cite{zenodo_learningtransitions}. Additionally, visualizations of the data, including an interactive version of the three-dimensional phase diagram of Fig.~\ref{fig:3DPhaseDiagram}, are provided there in the programming language Julia.

\appendix

\section{KL divergence and numerical discretization} 
\label{app:KL_divergence}

As given in the main manuscript the Kullback-Leibler divergence is defined as 
\begin{equation}
    D(P||Q) = \int P(\theta, \phi) \log \frac{P(\theta, \phi)}{Q(\theta, \phi)} d\theta d\phi \ ,
\end{equation}
where $P(\theta, \phi)$ is the probability distribution of the measurement outcomes on the Bloch sphere, and $Q(\theta, \phi)$ is the uniform distribution on the Bloch sphere. For numerical calculation of the probabilty distribution $P(\theta, \phi)$ from the points given by the calculated samples, we discretize the unit sphere into $N$ patches with (approximately) equal area $\Delta \Omega = 4\pi/N$. In this discretization the equal distribution over the sphere has the density $Q(\theta, \phi) = 1/N$ in each patch. The probability distribution $P(\theta, \phi)$ is estimated by counting the number of samples in each patch divided by the total number of samples and the area of the patch. With this discretization the Kullback-Leibler divergence can be approximated as
\begin{equation}
    D(P||Q) \approx \sum_{i=1}^N P_i \log \frac{P_i}{Q_i} = \sum_{i=1}^N P_i \log (P_i N)  
\end{equation} 
where $P_i$ is the estimated probability in patch $i$.

If the probability distribution $P(\theta, \phi)$ is uniform over the sphere, the estimated probability in each patch is $P_i = 1/N$ and the Kullback-Leibler divergence is zero. If the probability distribution $P(\theta, \phi)$ is a delta function at one point on the sphere, all samples fall into one patch and the estimated probability in that patch is $P_i = 1$ and zero in all other patches. In this case the Kullback-Leibler divergence sums to $\log(N)$. With a bimodal distribution with two delta functions at opposite points on the sphere, the Kullback-Leibler divergence sums correspondingly to $\log(N/2)$ and if the distribution is over  eight points in different patches to $\log(N/8)$.

\begin{figure}[t]
    \centering
    \includegraphics[width=\columnwidth]{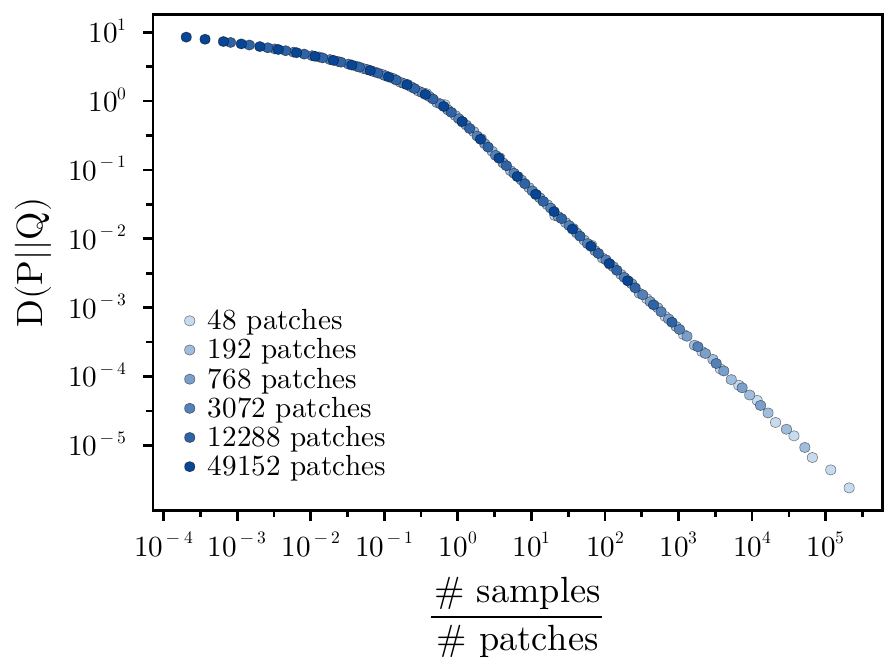}
    \caption{{\bf Kullback-Leibler divergence} of the uniform distribution with finite samples sampled according to the uniform distribution in a finite resolution. Plotted is the Kullback-Leibler divergence of the perfect uniform distribution with different resolutions of the HEALPix scheme (Hierarchical Equal Area isoLatitude Pixelation of a sphere) with $N=12 \cdot 4^n$ patches of equal size, where $n$ is the order of the discretization and $N$ the number of patches in the respective discretization. Due to the finite number of samples even the Kullback-Leibler divergence of points sampled according to the uniform distribution takes small finite values. Since all discretizations follow the same curve, the Kullback-Leibler divergence of the perfect uniform distribution only depends on the ratio of number of samples over number of patches. In the limit of infinite samples the Kullback-Leibler divergence always goes to zero as the points are sampled by the distribution with which they are compared.}
    \label{fig:KL_uniform_samples}
\end{figure}

The plots shown in this manuscript are calculated with $N=192$. Thus the maximal Kullback-Leibler divergence is $\log(192) \approx 5.26$. The bimodal distribution has a Kullback-Leibler divergence of $\log(96) \approx 4.56$ and the distribution over eight points has a Kullback-Leibler divergence of $\log(24) \approx 3.18$. The discretization into $N=192$ patches is a compromise between the resolution of the probability distribution and the number of samples needed to estimate the probability distribution in each patch, since for finite samples even the perfect uniform distribution will show some deviation from a Kullback-Leibler divergence of zero. This deviation can be estimated by calculating the Kullback-Leibler divergence for a perfect uniform distribution with finite samples. The accuracy of the calculation is only dependent on the number of samples over the number of bins: $\frac{\textrm{samples}}{N}$. 
This is shown in Fig.~\ref{fig:KL_uniform_samples}. For the calculations shown in this manuscript we used $10^4$ samples which gives a deviation of the Kullback-Leibler divergence from zero of about $0.01$ (currently not yet all points converged to the full number of samples, so slightly higher deviations of same order of magnitude possible).

When the Kullback-Leibler divergence is normalized to the value  of the bimodal distribution $\log(N/2)$, the value one corresponds to a bimodal distribution and the value zero to a uniform distribution. This normalized Kullback-Leibler divergence is used in Fig.~3 of the main manuscript.
In such a normalization the limit of the ring distribution in the X-Z plane corresponds to exactly $\frac{1}{2}$. Both bimodal and ring distribution diverge to infinity in that limit, since these distributions are clearly distinct and thus far away from the uniform distribution over the whole sphere.
That the ratio between these distributions becomes exactly $\frac{1}{2}$ can be understood how we discretize into patches and how this number of patches scales with each step. In the resolution used for the calculations here, there are 192 patches of (in very good approximation) equal area. The bimodal distribution concentrates on exactly two of these patches, while the ring distribution covers the 18 patches around the X-Z plane. Thus the expected value here would be $\frac{\log(192/18)}{\log(192/2)} \approx 0.51$, which is very close to the numerically observed value of $0.5$. Increasing the number of patches $N$ will increase the number of patches by a factor of four in every step, while the number of patches covered by the ring distribution increases by a factor of two.
If $n$ is the number of additional steps in discretization, we obtain
\begin{equation}
    K(\text{bimodal}) = \log(\frac{192}{2}  \cdot 4^n) \ ,
\end{equation}
for the bimodal distribution and
\begin{equation}
    K(\text{ring}) = \log(\frac{192 \cdot 4^n}{18 \cdot 2^n}) = \log(\frac{192}{18} \cdot 2^n) \ ,
\end{equation}
for the ring distribution. That means for the ratio in the limit following:
 \begin{eqnarray}
    \tilde{K} 
    & = & \lim \limits_{n \to \infty}\frac{\log(\frac{192}{18} \cdot 2^n)}{\log(\frac{192}{2} \cdot 4^n)}  \nonumber \\
    & = & \lim \limits_{n \to \infty}\frac{\log(2^n)}{\log(4^n)} = \frac{\log(2)}{\log(4)} = \frac{1}{2} \ .
\end{eqnarray}

\section{Detailed Analysis of non-Hermitian dynamics in the post-selection limit}
\label{app:nonhermitianfermion}

In this appendix, we provide more details on the spectrum of the post-selected Floquet operator in Eq.~(\ref{eqn:FloquetU}) following Ref.~\cite{LuGrover2021spacetimeduality}.
The Floquet operator in Eq.~(\ref{eqn:FloquetU}) can be written in a more symmetric form as
\begin{align}
   U_F &=  \exp\left({\frac{J_d+i\phi_d}{4} \sum_j X_j} e^{\frac{J+i\phi}{2} \sum_j Z_j Z_{j+1}} e^{\frac{J_d+i\phi_d}{4} \sum_j X_j}\right) \,.
\end{align}
Define $H_1 =  \sum_j Z_j Z_{j+1}$ and $H_0 = \sum_j X_j$. Under the Jordan-Wigner transformation
\begin{align}
    X_j &= 1- 2 c^\dagger_j c_j, \quad c^+_j = \left( \prod^{j-1}_{\ell = 1} X_\ell \right) \sigma^+_j, \quad c^-_j = \left( \prod^{j-1}_{\ell = 1} X_\ell \right) \sigma^-_j \,,
\end{align}
where $\sigma^\pm_j = \frac{1}{2} (Z_j \pm i Y_j)$, we have 
\begin{align}
    Z_j &= \sigma^+_j + \sigma^-_j = \left[ \prod^{j-1}_{\ell = 1} (1-2c^\dagger_\ell c_\ell) \right] (c_j + c^\dagger_j) \,.
\end{align}
As a result, 
\begin{align}
   H_1 &= \sum^L_{j=1} (c^\dagger_j - c_j)(c^\dagger_{j+1} + c_{j+1}) \,,
\end{align}
where we impose the anti-periodic boundary condition $c_{L+1} = -c_1$ for the even fermion parity sector $e^{i \pi N} = e^{i \pi \sum_j c^\dagger_j c_j} = \prod^L_{j=1} (1-2c^\dagger_j c_j) = 1$ and the periodic boundary condition $c_{L+1} = c_1$ for the odd fermion parity sector $e^{i \pi N} = -1$. Subsequently, Fourier transform of the complex fermion operator $c_j = \frac{1}{\sqrt{L}} \sum_k e^{i k j} c_k$ leads to
\begin{align}
    H_1 &= 2 \sum_{k>0} \begin{pmatrix}
        c^\dagger_k & c_{-k}
    \end{pmatrix} \begin{pmatrix}
        \cos{k} & i \sin{k} \\
        -i\sin{k} & -\cos{k}
    \end{pmatrix} \begin{pmatrix}
        c_k \\
        c^\dagger_{-k}
    \end{pmatrix} \, .
\end{align}
Similarly, we have
\begin{align}
    H_0 &=  \sum_j X_j =  \sum_j (1-2c^\dagger_j c_j) \nonumber \\
    &\sim -2 \sum_{k>0} (c^\dagger_k c_k + c^\dagger_{-k} c_{-k}) \nonumber \\
    &=  -2 \sum_{k>0} \begin{pmatrix}
        c^\dagger_k & c_{-k}
    \end{pmatrix} \begin{pmatrix}
        1 & 0 \\
        0 & -1
    \end{pmatrix} \begin{pmatrix}
        c_k \\
        c^\dagger_{-k}
    \end{pmatrix} \,,
\end{align}
up to a complex constant. As a result, the Floquet operator can be written in the momentum space as
\begin{equation}
    \begin{split}
        U_F &\propto \prod_{k>0} e^{-\frac{J_d + i \phi_d}{2} \sigma_z} \cdot e^{(J+i\phi) \sigma_k} \cdot e^{-\frac{J_d + i \phi_d}{2} \sigma_z} \,,
    \end{split}
\end{equation}
where $\sigma_k = \sigma_z \cos{k} - \sigma_y \sin{k}$. 
In particular, at Y point of the phase diagram, we have $J = J_d = 0, \phi = -\phi_d = \pi/2$. Consequently, the Floquet operator reduces to
\begin{equation}
\begin{split}
    U_F^Y &= \exp\left({\frac{i\pi}{4} \sum_j Z_j Z_{j+1}} e^{-\frac{i\pi}{4} \sum_j X_j}\right) \\
    & = \prod_{k>0} e^{i \frac{\pi}{2} \sigma_k} e^{i \frac{\pi}{2} \sigma_z} = \prod_{k>0} (- e^{-ik \sigma_x}) \,,
    \end{split}
\end{equation}
which is a {\it unitary} operator. 

\bibliography{measurements}

\end{document}